\documentclass[print]{aa}

\usepackage{times}
\usepackage{caption}
\usepackage[round]{natbib}
\usepackage{pdflscape}
\usepackage{afterpage}
\usepackage{capt-of}
\usepackage{color}
\usepackage{hyperref}
\usepackage{graphicx}
\usepackage{longtable}
\usepackage{tabu}

\usepackage{multirow}
\usepackage{amsmath}
\usepackage{verbatim}
\usepackage{pdflscape}
\usepackage{caption}
\usepackage{subcaption}
\usepackage{amsmath}
\usepackage[normalem]{ulem}
\usepackage[autostyle]{csquotes}

\begin{document}

\title{Non-Vestoid candidate asteroids in the inner main belt}

\authorrunning{Oszkiewicz et al.}

\titlerunning{Non-Vestoid candidate asteroids in the inner Main Belt}

\author{Dagmara Oszkiewicz$^{1,2}$ \and Brian A. Skiff$^2$ \and Nick Moskovitz$^2$ \and Pawe{\l} Kankiewicz$^3$ \and Anna Marciniak$^2$ \and Javier Licandro$^{4,5}$ \and Mattia Galiazzo$^6$ \and Werner Zeilinger$^7$}

 \institute{Astronomical Observatory Institute, Faculty of Physics, A. Mickiewicz University, S{\l}oneczna 36, 60-286 Pozna{\'n}, Poland \and
Lowell Observatory, 14000 W Mars Hill Rd, 86001 Flagstaff, AZ, USA \and
              Institute of Physics, Astrophysics Division, Jan Kochanowski University, Swietokrzyska 15, 25-406 Kielce, Poland \and
              Instituto de Astrof\'{\i}sica de Canarias (IAC), C\/V\'{\i}a L\'{a}ctea s\/n, 38205 La Laguna, Spain \and
              Departamento de Astrof\'{\i}sica, Universidad de La Laguna, 38206 La Laguna, Tenerife, Spain \and
              The Department of Physics and Astronomy, University of Western Ontario, 1151 Richmond Street, London, Ontario, Canada \and
              Institut f\"ur Astrophy der Universit\"at Wien, T\"urkenschanzstra\ss e 17, A-1180 Wien, Austria         }

\date{\today} 

 \offprints{D. Oszkiewicz, \\ e-mail: {\tt oszkiewi@astro.amu.edu.pl}}

   \date{Received xx xx xxxx / Accepted xx xx xxxx} 
   
 \abstract
 {{\bf{Most howardite-eucrite-diogenite (HED) meteorites (analogues to V-type asteroids) are thought to originate from the asteroid (4) Vesta. However some HEDs show distinct oxygen isotope ratios and therefore are thought to originate from other asteroids. In this study we try to identify asteroids that may represent parent bodies of those mismatching HEDs. 
 }}}
 {{\bf{The main goal of this study is to test the hypothesis that there might be V-type asteroids in the inner main asteroid belt unrelated to (4) Vesta. In order to evolve outside the Vesta family and became Vesta fugitives, asteroids should produce the correct Yarkovsky drift. The direction of which is dependent on asteroid sense of rotation. Therefore we focus on determining sense of rotation for asteroids outside the Vesta family to better understand their origin.
  }}}
 {{\bf{We performed photometric observations using the 1.1 m and 1.8 m telescopes at Lowell Observatory to determine rotational synodic periods of selected objects before, at, and after opposition. Prograde rotators show a minimum in synodic period at opposition while retrograde rotators show a maximum. This is known as the "drifting minima" method. Changes in the rotational period are on the order of seconds and fractions of seconds and depend on the rotational pole of the object and the asteroid-observer-Sun geometry at opposition.
 }}}
{{\bf{We have determined sense of rotation for eight asteroids and retrieved spin states for three objects from literature. For one asteroid we were not able to determine the sense of rotation. In total our sample includes 11 V-type asteroids and one S-type (test object). We have revised rotation periods for three objects. Five V-types in our sample can be explained by migration from the Vesta family. Two show spin states that are inconsistent with migration from Vesta. The origin of the remaining objects is ambiguous. 
}}}
{{\bf{We found two objects with rotations inconsistent with migration from Vesta. Assuming that the YORP effect and random collisions did not substantially modify their sense of rotation, those objects are candidates for non-Vestoids in the inner asteroid belt. Finding more non-Vestoids is crucial in testing the formation and migration theory of differentiated parent bodies. 
}}}



\maketitle

\section{Introduction}
The \enquote{missing mantle} problem is a long standing puzzle in planetary science. While the meteorite record shows extensive evidence for perhaps 100 differentiated parent bodies (i.e., bodies segregated into an iron core, silicate mantle, and basaltic crust) in the early Solar System, there has yet to be found an equivalent record amongst asteroids. In particular there appears to be a lack of mantle and basaltic crustal fragments in the main belt, hence the missing mantle problem \citep{burbine1996mantle}. 

Most basaltic (V-type) asteroids and analogous howardite-eucrite-diogenite (HED) meteorites can be linked to a single asteroid - (4) Vesta. That connection has been known for almost half a century now. Spectroscopic observations of Vesta, HED meteorites and asteroids in the vicinity of Vesta show the same spectral characteristics \citep{mccord1970asteroid,binzel1993chips}. The distribution of cosmic ray exposure ages among HEDs shows two peaks interpreted as major cratering events on Vesta that produced the majority of those meteorites. Indeed the two large craters were identified on the surface of Vesta (\cite{thomas1997impact}, \cite{2012Schenk}) making the cratering and Earth delivery scenario plausible. Furthermore, the distribution of V-type asteroids in the inner main belt stretches from the Vesta family asteroids toward major resonances making Earth delivery possible. The combination of the Yarkovsky thermal effect and dynamical interactions with resonances creates an efficient delivery mechanism (\cite{scholl1991nu} \cite{Vokrouhlicky1998} \cite{Vokrouhlicky1999}).

For these reasons V-type asteroids in the inner main belt are generally believed to be collisionally freed fragments of Vesta. However, as highlighted by the fall of the anomalous HED meteorite Bunburra Rockhole, there might be basaltic asteroids in the inner main asteroid belt that are not genetically linked to (4) Vesta and may represent other differentiated parent bodies \citep{bland2009anomalous}. The oxygen isotope ratios of the Bunburra Rockhole meteorite are distinct from those of typical HEDs, therefore ruling out its Vestoidal origin. The fall of the meteorite was observed and its orbit was traced back to the inner main asteroid belt. An origin in the mid- or outer-main belt was excluded. The parent body of this meteorite could still be present in the inner main belt. Other studies have also shown that there may be non-Vestoid V-types in the inner main belt. For example asteroid (809) Lundia previously considered a Vesta fugitive has a prograde rotation and small positive Yarkovsky drift, bringing into question its proposed evolutionary path \citep{carruba2005v} and origin in the Vesta family \citep{oszkiewicz2015differentiation}. Another asteroid, (2579) Spartacus, located in the inner main belt shows spectral characteristics slightly different those of typical Vestoids (shifted band centers and deeper absorption bands \cite{burbine2001vesta}, \cite{ieva2016spectral}).

A handful of V-type asteroids have been found in the mid- and outer-main asteroid belt ( e.g., \cite{lazzaro2000discovery}, \cite{solontoi2012avast}, \cite{duffard2009two}, \cite{moskovitz2008spectroscopically}, \cite{roig2008v} and others). Due to the separation by the strong 3:1 resonance with Jupiter those cannot be easily connected with (4) Vesta. The spectral properties of those objects also differ from typical Vestoids (\cite{ieva2016spectral}) strengthening the idea that those objects represent other differentiated parent bodies. Possible scenarios explaining the scattered distribution of those mid- and outer-main belt objects have been proposed. Through dynamical integration \cite{Carruba2014} showed that most low inclination V-types in the mid main belt could evolve from the parent bodies of the Eunomia and Merxia/Agnia families on time-scales of 2 Gyr. Higher inclination V-types were not reproduced. Another scenario was proposed by \cite{bottke2006iron} and revised in \cite{bottke2014origin}. He supposes that the differentiated parent bodies may have formed in the terrestrial planet region 4-6 Gyr ago and were either gravitationally scattered into the main belt by planetary embryos and/or scattered and captured due to planet migration in the Grand Tack model. The number of differentiated asteroids in this scenario varies from one to several hundred depending on the formation region and minimum body diameter required for full differentiation. Both scenarios can partially explain the distribution of the observed V-types in the mid- and outer-main belt. Identification of new non-Vestoids in uncharted territories like the inner main belt may help further constrain those scenarios as well as the number, formation location and evolution of differentiated parent bodies.

The main goal of this study is therefore to test whether there are asteroids in the inner main belt that cannot be linked to (4) Vesta and if those may represent parent bodies of meteorites such as Bunburra Rockhole. This is a challenging task and it has not been previously attempted for several reasons. First the inner main belt is interwoven with multiple dynamical resonances making the studies of dynamical history of individual objects and populations in the region very challenging. Second, the inner main belt contains multiple overlying asteroid families making the family membership of many asteroids in the region ambigious. Finally, the proximity of the asteroid (4) Vesta and its collisional family, as mentioned earlier, led to the conclusion that most V-types in this region are fragments of asteroid (4) Vesta that were freed from its surface in violent collisions and then migrated to the inner parts of the asteroid belt. 

To face those challenges we propose to combine several approaches. First we will use photometric observations to determine sense of rotation of asteroids in the inner main belt outside the dynamical Vesta family. The sense of rotation determines the direction of the Yarkovsky drift - the dominant scattering mechanism for asteroids smaller then 40 km in the inner main belt (\cite{2012Delisle}). Retrograde rotators are driven inwards toward the Sun, while prograde rotators migrate outwards. The direction of the Yarkovsky drift is therefore crucial when considering whether or not an object escaped from the Vesta family (\cite{nesvorny2008fugitives}, \cite{2012Delisle}). \cite{nesvorny2008fugitives} simulated the escape paths from Vesta and its family, showing that typical Vesta fugitives in the inner main asteroid belt at semi-major axis a $< 2.3$ AU must have retrograde rotation and physical and thermal parameters that maximize the Yarkovsky force in order to evolve to scattered orbits within 1-2 Gys (age of the Vesta collisional family \cite{spoto2015asteroid}, \cite{carruba2016constraints}). Therefore large asteroids outside the Vesta family with a $< 2.3$ AU and having thermal and rotational properties minimizing the Yarkovsky drift or showing Yarkovsky drift direction toward (4) Vesta are the best candidates for non-Vestoidal V-type asteroids in this region. Throughout this work we neglected random collisions and so-called Yarkovsky-O'Keefe-Radzievskii-Paddack (YORP) effect. In particular large and nearly spherical objects are less susceptible to spin modification by the YORP effect. 

In this study we focus on determining the spin properties of inner belt V-type asteroids to help better understand the dynamical evolution of those objects and whether they originated from Vesta or some other body. In further studies we will focus on the spectroscopic and dynamical properties of those objects. In section \ref{methods} we describe the target selection process, photometric observations and the drifting minima method used for determination of the direction of rotation. In section \ref{res} we present the results, in section \ref{Yarkovsky} we estimate the maximum Yarkovsky drift and YORP timescale and in section \ref{concl} we summarize our conclusions and discuss future work.

\section{Methods}
\label{methods}

\subsection{Target selection}
For this survey we have selected a number of V-type asteroids outside the Vesta dynamical family, at various locations in the inner main asteroid belt. The V-type candidates were selected based on Sloan Digital Sky Survey (SDSS) colors (\cite{ivezic2001solar}) using the method described in \cite{oszkiewicz2014}. For purely practical reasons (minimize observing time needed) we focus on asteroids with rotational periods P$<$ 12h and objects for which some previous data is available in the Minor Planet Center Lightcurve database (\cite{warner2009asteroid}). For completeness we also include known V-types outside the Vesta dynamical family whose rotational properties are known (some of which have periods longer then P$>$ 12h). The target asteroids along with their absolute magnitudes, osculating elements (a, e, i) and other parameters are listed in Table \ref{targets}. 

For reference we computed ejection velocity from Vesta for each target based on the proper elements as (\cite{carruba2007v}):

\begin{equation}
\Delta v = n  \cdot a  \cdot \sqrt{1.25 \cdot \bigg{(}\frac{\Delta a}{a}\bigg{)}^2+2 \cdot (\Delta e)^2 + 2 \cdot (\Delta sin (i))^2},
\end{equation}

where $n$ is the proper mean-motion and $\Delta v$ is the velocity distance between an asteroid and (4) Vesta. Proper elements ($a$ - proper semi-major axis, $e$ - proper eccentricity, $i$ - proper inclination) were taken from the AstDys database \footnote{\url{http://hamilton.dm.unipi.it/astdys/}}. The distances in proper elements space between Vesta and an asteroid are denoted as ${\Delta a}$, ${\Delta e}$, and ${\Delta sin(i)}$.  \cite{asphaug1997impact} estimated that multikilometer fragments can be ejected from Vesta with terminal speeds of $\Delta v \sim$ 600 $m \cdot s^{-1}$ or less.

Where available we list the asteroid diameters (\cite{masiero2011main}, \cite{mainzer2011preliminary}, \cite{nugent2016neowise}). Otherwise the diameters were computed as $D = \frac{1329}{\sqrt{p}}10^{-0.2H}$ assuming average V-type albedo ($p_V=0.293 \pm 0.150$, \cite{usui2012albedo}) - denoted with a star in table \ref{targets}.

For comparison to \cite{nesvorny2008fugitives} we note in Table \ref{targets} which asteroids are located in kinematic cells defined in that work. According to \citep{nesvorny2008fugitives} fugitives from Vesta that evolve to Cell I (defined in terms of orbital parameters as 2.2 AU $<$ a $<$ 2.3 AU, 0.05 $<$ e $<$ 0.2, 0$^{\circ}$ $<$ i $<$10$^{\circ}$) typically should have retrograde rotations and thermal parameters that maximize Yarkovsky drift. Objects that are prograde rotators in this region are less likely to migrate from Vesta. Cell II corresponds to low inclination objects (2.32 AU $<$ a $<$ 2.48 AU, 0.05 $<$ e $<$ 0.2, 2$^{\circ}$ $<$ i $<$6$^{\circ}$). \cite{nesvorny2008fugitives} were not able to reproduce this population by migration from the Vesta family over 1-2 Gy integration with sufficient efficiency. Therefore they proposed that those objects are either (1) fragments of differentiated bodies other than (4) Vesta, or (2) they were freed from the surface of (4) Vesta before or during the heavy bombardment epoch ($\sim$ 3.8 Gy ago), and then evolved dynamically with the help of secular resonances sweeping though the region.

Taking into consideration those results, the best candidates for non-Vestoids in the inner main belt would be large asteroids outside the dynamical Vesta family having large $\Delta v$ values and rotational properties indicative of current migration direction toward (4) Vesta.

\subsection{Photometric observations and data reduction}

 Photometric observations were conducted using the several
telescopes listed in Table \ref{telescopes}.  Most of the data were taken with
a Cousins R filter, but some with Johnson V, or a wideband VR
filter.  Altogether we obtained data on 105 nights, mostly at
Lowell Observatory.  We performed standard photometric reductions
with bias and flat-field correction followed by ordinary aperture
photometry.  We used the commercial software {\it Canopus}
(version 10.4.0.6), making use of its built-in photometric
catalog to obtain approximate Cousins R or Sloan {\it r'} nightly
zero-points.  We generally used four or five comparison stars in
each field, usually of much higher signal-to-noise than the target,
and selected to have near-asteroidal colors if possible.
For asteroids near their stationary points, we could use the
same comparison star sets on multiple nights.  The measuring
apertures were generally 10 to 15 arcsec diameter depending on
image quality.  The typical mean nightly {\it rms} scatter of
the ensemble of comparison stars was 0.007 mag.

The rotational periods were computed using standard Fourier series fitting and the period uncertainties were estimated using a direct Monte Carlo simulation of the photometric uncertainties. That is, for each asteroid we generated 100 000 photometric datasets sampled within the photometric uncertainties of the nominal measurements. Each dataset was used to estimate the period using the standard Fourier series resulting in a sample of possible periods for each asteroid. The period uncertainty was then computed as  99.73002\% probability mass (equivalent of 3-$\sigma$ uncertainty). The photometric uncertainties were conservatively assumed to be around $\sim$0.05 mag, even though in most cases the photometry was more accurate. The order of the Fourier series was adjusted as the lowest order that reasonably fit the data.

\begin{table}[htbp]
\begin{center}
\begin{tabular}{|lllll|} \hline
Telescope       & Observatory                           & Instrument            & Aperture        & Nights \\ \hline
Perkins         & Lowell Observatory            & prism                 & 1.8 m           & 47 \\
Hall                    & Lowell Observatory            & nasa42                        & 1.1 m           & 46 \\
31 inch         & Lowell Observatory            & nasa31                        & 0.7 m           & 16 \\
IAC-80          & Teide Observatory                     & camelot                       & 0.8 m           & 3 \\  
\hline
Total:                  &                                               &                               &                       & 112 \\
\hline
\end{tabular}
\caption{Telescopes and instruments used.}
\label{telescopes}
\end{center}
\end{table}

\subsection{The drifting minima method}
We determined senses of rotation by measuring synodic periods of the selected objects before, during and after opposition. This allows us to see the changes in synodic period as the asteroid moves toward and away from opposition. Prograde rotators have their synodic period increasing when they move away from opposition (minimum at opposition) while retrograde rotators have a maximum synodic period at opposition. The amount of the change in synodic period depends on orientation of the rotational pole. 
Changes in synodic period are typically very small (fraction of a second). A high precision and accuracy of period estimation is required to detect these changes. Such precision can only be achieved with high quality photometric lightcurves. 
This method, often referred to as the drifting minima method, is well established in the literature; we refer the reader to \cite{groeneveld1954photometric} \cite{gehrels1967minor} \cite{magnusson1989pole} and \cite{dykhuis2016efficient} for more details.

In general, the relationship between the synodic (with respect to the observer) and sideral (with respect to the stars) periods can be given in terms of change in the ecliptic longitude of the phase angle bisector (PAB) $l_p$ (\cite{dykhuis2016efficient} \cite{magnusson1989pole}):
 
\begin{equation}
P_{syn}^{-1} = P_{sid}^{-1} - \frac{\Delta l'}{\Delta t},
\end{equation}

where $l'$ is the PAB measured relative to the north rotation pole of the asteroid, $P_{syn}$ is the synodic period, and $P_{sid}$ is sideral period. For retrograde rotators $\frac{\Delta l'}{\Delta t} < 0$ and for prograde rotators $\frac{\Delta l'}{\Delta t} > 0$. In other words, objects that show minimum in the synodic period at opposition have a prograde spin sense, while objects that show a maximum have a retrograde sense of rotation. We note that only the sense of rotation can be reliably estimated with this method. The spin coordinates estimated with the method are approximate and burdened with large uncertainties.

The key advantage of this method is the fact that the sense of rotation can be obtained with data from a single opposition. However the uncertanity on the exact pole coordinates is large, therefore the method can only be used to determine sense of rotation and not the detailed pole coordinates.
On the other hand methods relying on lightcurve inversion techniques require data from a minimum of four oppositions (usually around five to six years to observe the objects at various aspects), but the exact pole coordinates and shape of the object in addition to period and sense of rotation can be found. However for the purpose of this study the detailed pole orientation is not needed as the direction of the Yarkovsky drift is determined by the sense of rotation and therefore the drifting minima method is sufficient.

\section{Sense of rotation}
\label{res}
We have determined sense of rotation using the drifting minima method for eight asteroids, and for three asteroids sense of rotation was known from the literature. Five objects are located in cell I, two in cell II. The remaining ones outside the cells. We found five prograde rotators and three retrograde rotators. For three asteroids we have revised their rotational period. From the literature we added two retrograde rotators and one prograde rotator. The results are summarized in Table \ref{targets}. 

We found five objects whose rotational properties are consistent with origin in the Vesta family. The remaining objects are either from another differentiated parent body, or may have a complicated dynamical history possibly including spin flipping due to the YORP effect or collisions. Below we discuss each asteroid separately.

\subsection{(222) Lucia}
\begin{figure}[htbp]
   \centering
   \resizebox{0.5\hsize}{!}{\includegraphics[width=0.5 \textwidth]{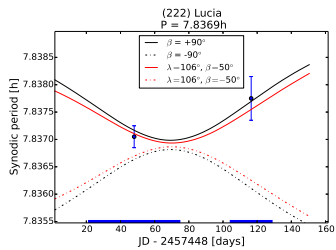}} 
   \caption{Modeled synodic period for asteroid (222) Lucia. Models for various spin-axis orientations are shown with solid lines (prograde models) and dashed lines (retrograde models). Points are the measured synodic periods with uncertainties. Thick lines at the bottom of the plot indicate the time-span over which the observations were taken.}
   \label{Lucia_mod}
\end{figure}

Asteroid (222) Lucia is an S-type object which we used to confirm that our period determination is sufficient to reproduce known results. Asteroid (222) Lucia is a prograde rotator with a known rotational pole: $\alpha \sim 106^{\circ}, \beta \sim 50 ^{\circ}$ (\cite{hanuvs2015new}). In Figs. \ref{LuciaOpp} and \ref{LucisAfter} we show lightcurves measured around and after opposition. The synodic period at opposition was $P_{at} = 7.8371 \pm 0.0002$ h and after $P_{after} = 7.8378 \pm 0.0004$ h. The increase in the synodic period is consistent with prograde rotation. In Fig. \ref{Lucia_mod} we show the modeled change in synodic period for spin axis longitude $\beta=90^{\circ}$ and $\beta=-90^{\circ}$, the known prograde spin orientation ($\alpha = 106^{\circ}, \beta = 50 ^{\circ}$), and the symmetric retrograde solution ($\alpha = 106^{\circ}, \beta = -50 ^{\circ}$). As expected the measurements fit well to the model derived with the pole coordinates from \cite{hanuvs2015new}. The drifting minima method is therefore reliable and can be used to determine sense of rotation.

\subsection{(809) Lundia}
We did not observe (809) Lundia in this study. We include this object here for completeness. Asteroid (809) Lundia is a V-type object with spectral characteristics indistinguishable from typical Vestoids (\cite{moskovitz2010spectroscopic}, \cite{Birlan2014}). Lundia has been previously considered a Vesta fugitive \citep{carruba2005v}. \cite{carruba2005v} showed that with a combination of dynamical resonances and the Yarkovsky effect a retrograde rotating object could migrate from within the Vesta family to the current location of Lundia. However from previous observations we know that (809) Lundia has a prograde rotation \citep{kryszczynska2009new}, is large (equivalent diameter 9.1 km) and is located well outside the Vesta family. Therefore the evolutionary path proposed for this object by \cite{carruba2005v} in unlikely. \cite{oszkiewicz2015differentiation} showed through dynamical integration that (809) Lundia has a very small Yarkovsky drift and the direction of migration is inconsistent with migration from Vesta. The YORP effect is unlikely to significantly modify the spin of a body of Lundia's size. It should however be noted that (809) Lundia is a binary asteroid and therefore BYORP may play a role (e.g., \cite{steinberg2011binary}).

\subsection{(1946) Walraven}
We did not observe (1946) Walraven for this study. However the object has a known spin orientation (\cite{hanuvs2015new}, \cite{durech2016asteroid}) and is in retrograde rotation ($\lambda=259^{\circ}$, $\beta=-80^{\circ}$ or $\lambda=20^{\circ}$, $\beta=-59^{\circ}$). We include it here for completeness. The object is located in Cell I and its retrograde rotation is consistent with an origin from within Vesta family.

\subsection{(2704) Julian Loewe} 
\begin{figure}[htbp]
   \centering
   \resizebox{0.5\hsize}{!}{\includegraphics[width=0.5 \textwidth]{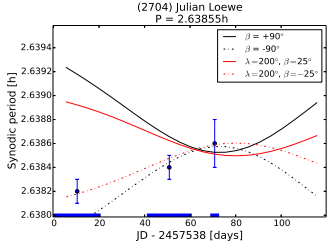}} 
   \caption{Model synodic period as in Fig.\ref{Lucia_mod}, but for (2704) Julian Loewe.}
   \label{Loewe_mod}
\end{figure}

Asteroid (2704) Julian Loewe is a spectroscopically confirmed V-type object (\cite{bus2002phase}). We observed Julian Loewe at four epochs: twice before opposition, once at opposition (date of opposition: 2016/08/09) and once after the opposition. Composite lightcurves (see Fig. \ref{loewe}) show four maxima indicating a complex shape. The measured periods before the opposition were $P_{before_1}=2.6382\pm 0.0001$h and $P_{before_2}=2.6383\pm 0.0001$h and at opposition $P_{at}=2.6386\pm 0.0002$h and after opposition $P_{at}=2.6383\pm 0.0002$h. Based on our observations across multiple rotations and multiple nights we revise the rotational period from $P_{syn}=2.091$h (\cite{Clark2015}) to $P_{syn}=2.6383\pm 0.0001$h. The data at opposition are dense and good quality, therefore in estimating the period uncertainty we assumed less conservative, more realistic 0.02 mag photometric uncertainty. The change in the synodic period is positive, thus the data are best explained by retrograde rotation. However it should be noted that the epoch method is less reliable for low pole asteroids. The model is shown in Fig. \ref{Loewe_mod}. Asteroid (2704) Julian Loewe belongs to the low inclination population and its origin is ambiguous at this point. It either is unrelated to Vesta or (because of its retrograde rotation, negative Yarkovsky drift, and current semi-major axis) started at large semi-major axis in the Vesta family and migrated to its current position. More detailed dynamical studies are required to fully comprehend the dynamical history of this object.

\subsection{(4796) Lewis}
\begin{figure}[htbp]
   \centering
   \resizebox{0.5\hsize}{!}{\includegraphics[width=0.5 \textwidth]{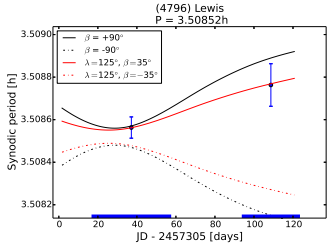}} 
   \caption{Model synodic period as in Fig.\ref{Lucia_mod}, but for (4796) Lewis.}
   \label{Lewis_mod}
\end{figure}

Asteroid (4796) Lewis is a confirmed V-type object with spectral characteristics indistinguishable from typical Vestoids (\cite{bus2002phase}, \cite{moskovitz2010spectroscopic}). We observed Lewis at two epochs: at opposition and after opposition (date of opposition: 2015/11/07). See the respective composite lightcurves in Figs. \ref{LewisOpp} and \ref{LewisAfter}. The measured synodic period during the opposition was $P_{at}=3.5086$ h $\pm 0.00005$ h and after the opposition $P_{after}=3.5088$ h $\pm 0.0001$ h. 
The synodic period is increasing and therefore consistent with prograde rotation. The modeled synodic period is shown in Fig. \ref{Lewis_mod}.
Prograde rotation implies positive values of the Yarkovsky drift ($\frac{da}{dt} > 0$). 
Like (2704) Julian Loewe, asteroid (4796) Lewis is in the low inclination population (Cell II) and its origin is ambiguous at this point. It is either unrelated to Vesta or (because of its prograde rotation, positive Yarkovsky drift, and current semi-major axis) it started at small semi-major axis in the Vesta family and migrated to its current position. More detailed dynamical studies are required to fully comprehend the dynamical history of this object.

\subsection{(5150) Fellini}
\begin{figure}[htbp]
   \centering
   \resizebox{0.5\hsize}{!}{\includegraphics[width=0.5 \textwidth]{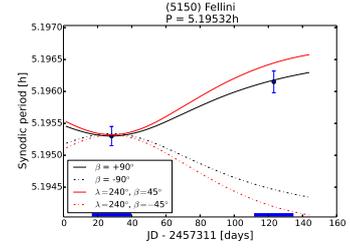}}
   \caption{Model synodic period as in Fig.\ref{Lucia_mod}, but for (5150) Fellini.}
   \label{Fellini_mod}
\end{figure}

We observed the asteroid (5150) at two epochs: at opposition and after opposition (date of opposition: 2015/11/12). The composite lightcurves are shown in Fig. \ref{FelliniOpp} and \ref{FelliniAfter}. The synodic period measured at opposition was $P_{at} = 5.1953$h $\pm 0.0002$h and after $P_{after} = 5.1962$h $\pm 0.0002$h. The increase in the synodic period is consistent with prograde rotation. The modeled change in synodic period is shown in Fig. \ref{Fellini_mod}. The data are best explained by high rotational pole ($\beta$ close to 90 degrees). The Yarkovsky drift for this asteroid might therefore be maximized. The asteroid is located at semi-major axis larger than most Vesta family objects. The prograde rotation and positive Yarkovsky drift is in line with migration from the Vesta family. Thus (5150) Fellini is most likely a fugitive from Vesta. The estimated maximum Yarkovsky drift for this asteroid is $da/dt=9.047e-5$ AU/Myr giving approximate migration time from Vesta $<$ 1.4 Gyr. This result is consistent with the known age of the Vesta family ($\sim$ 1-2Gyr).

\subsection{(5525) 1991 TS4}

Asteroid (5525) 1991 TS4 has a preliminary period estimation $P=6.972$h \footnote{\url{http://obswww.unige.ch/~behrend/}}.
We observed the object and found that its period is more than twice as long as this estimation, that is $P_{syn} = 14.088 \pm 0.004$h. Due to a combination
of bad weather, long period and telescope scheduling we were not able to obtain a reliable second epoch for this object. Therefore we are not able to determine its
sense of rotation. We are including it here only because of the revised rotational period. The composite lightcurves for (5525) 1991TS4 are shown in Fig. \ref{5525_lc}.
Formally the measured periods were: $P_{before} = 13.9 \pm 0.2$h, $P_{at} = 14.088 \pm 0.004$h, $P_{after} = 14.08 \pm 0.01$h, however other solutions 
could not be excluded.

\subsection{(5599) 1999 SG1}
\begin{figure}[htbp]
   \centering
   \resizebox{0.5\hsize}{!}{\includegraphics[width=0.5 \textwidth]{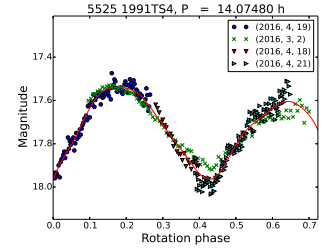}}
   \caption{Model synodic period as in Fig.\ref{Lucia_mod}, but for (5599) 1999 SG1.}
   \label{5599_mod}
\end{figure}

We observed asteroid (5599) 1999 SG1 at two epochs: at opposition and after opposition (date of opposition: 2016/05/25). Composite lightcurves are shown in fig. \ref{5599lc}.
The measured period at opposition was $P_{at} = 3.6195 \pm  0.0001$h and after  $P_{after} = 3.6201 \pm  0.0001$h. 
These data are consistent with prograde rotation (the model is shown in Fig. \ref{5599_mod}) and thus (5599) 1999 SG1
has a positive Yarkovsky drift. The asteroid is located at large semi-major axis compared to Vesta, and therefore can be explained
by migration from the Vesta family.

\subsection{(5754) 1992 FR2}
\begin{figure}[htbp]
   \centering
   \resizebox{0.5\hsize}{!}{\includegraphics[width=0.5 \textwidth]{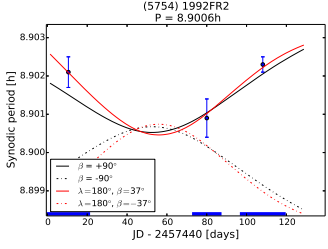}} 
   \caption{Model synodic period as in Fig.\ref{Lucia_mod}, but for (5754) 1992 FR2.}
   \label{5754_mod}
\end{figure}

We observed the asteroid (5754) 1992 FR2 at three epochs: before, at and after opposition (date of opposition: 2016/04/17). Composite lightcurves are shown in Fig. \ref{5754lc}. The synodic period measured before opposition was
$P_{before}=8.9021$h $\pm0.0004$h at opposition $P_{at}=8.9010$h $\pm0.0005$h and after $P_{after}=8.9024$h $\pm 0.0005$h. The change in synodic period is consistent with prograde rotation (the modeled change in synodic period is shown in Fig. \ref{5754_mod}), thus 
a positive Yarkovsky drift ($\frac{da}{dt} > 0$). We estimate the maximum Yarkovsky drift for this object $\frac{da}{dt}_{max}=1.864\times10^{-4}$ AU/My. Asteroid (5754) 1992 FR2 is located in Cell I and has a positive Yarkovsky drift
thus is unlikely to be a Vesta fugitive. Therefore asteroid (5754) 1992 FR2 is a non-Vestoid candidate.

\subsection{(5875) Kuga}
\begin{figure}[htbp]
   \centering
   \resizebox{0.5\hsize}{!}{\includegraphics[width=0.5 \textwidth]{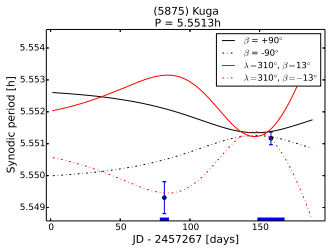}}
   \caption{Model synodic period as in Fig.\ref{Lucia_mod}, but for (5875) Kuga.}
   \label{Kuga_mod}
\end{figure}

We have observed asteroid (5875) Kuga at two epochs: before and at opposition (date of opposition: 2016/01/26). Composite lightcurves are shown in Fig. \ref{5857lc}. The synodic period measured before opposition was
$P_{before}=5.5493$h $\pm0.0005$h and at opposition $P_{at}=5.5512$h $\pm0.0002$h. The increase in the synodic period is consistent with retrograde rotation (the modeled change in synodic period is shown in Fig. \ref{Kuga_mod})
implying negative Yarkovsky drift ($\frac{da}{dt} < 0$). Asteroid (5875) Kuga is outside the Vesta family and has a negative Yarkovsky drift ($da/dt \sim -7.458\times10^{-5}$ AU/Myr),
which means that it must have originated at even larger semi-major axis. The origin of this body in ambiguous at this point. It is ether unrelated to Vesta or (because of its retrograde rotation, negative Yarkovsky drift, and current semi-major axis) started 
at large semi-major axis in the Vesta family and migrated to its current position. More detailed dynamical studies are required to fully comprehend the dynamical history of this object.

\subsection{(6406) 1992 MJ}

We did not observe (6406) 1992 MJ in the course of this study. However the object has a known spin orientation from \cite{hanuvs2015new}, and \cite{durech2016asteroid} and is retrograde rotating ($\lambda=17^{\circ}$, $\beta=-61^{\circ}$ or $\lambda=216^{\circ}$, $\beta=-52^{\circ}$). We include it here for completeness. The object is located in Cell I and its retrograde rotation is consistent with origin in the Vesta family.

\subsection{(18641) 1998 EG10}
\begin{figure}[htbp]
   \centering
   \resizebox{0.5\hsize}{!}{\includegraphics[width=0.5 \textwidth]{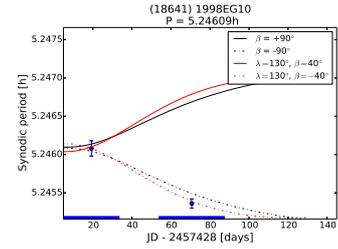}} 
   \caption{Modeled synodic period.}
   \label{18641_mod}
\end{figure}

We observed (18641) 1998 EG10 at two epochs: at and after opposition. Composite lightcurves are shown in Fig. \ref{18641lc}. We have revised the rotational period from the previous estimation $P=5.68$h (\cite{Clark2008}) to $P=5.2461$h $\pm 0.0001$h. The measured period at opposition was $P_{at}=5.2461$h $\pm 0.0001$h and after opposition $P_{after}=5.2454$h $\pm 0.0001$h. The decreasing synodic period is consistent with retrograde rotation thus implying negative Yarkovsky drift. The modeled change in synodic period is shown in Fig. \ref{18641_mod}. Given its location and negative Yarkovsky drift (18641) 1998 EG10 is most likely a Vesta fugitive. It should also be noted that (18641) 1998 EG10 has a small $\Delta v$ and thus by extending the $\Delta v$ limit in family clustering algorithms, this object could also easily be attributed to the Vesta family.

\section{Dynamical and spin modification effects}

\subsection{Estimating Yarkovsky drift}
\label{Yarkovsky}
For reference we have estimated maximum possible Yarkovsky drift rates for our targets given their diameters and the determined direction of rotation (Table \ref{Yarko}). The maximum possible solar radiation drift was computed as (\cite{Nugent2012}):

\begin{equation}
\frac{da}{dt}=f_Y\frac{3}{4\pi}\frac{1}{\sqrt{1-{e^2}}}\frac{L_{\odot}}{GM_{\odot}}\frac{1}{D\rho},
\end{equation}

or in convenient units:

\begin{equation}
\begin{aligned}
\frac{da}{dt} = & \frac{1.457}{\sqrt{1-{e^2}}}\left(\frac{f_Y}{10^{-5}}\right)\left(\frac{1 \; km}{D}\right)\times \\
                        & \left(\frac{1000 \; kg \; m^{-3}}{\rho}\right)\; 10^{-3} \; AU \; Myr^{-1}.
\end{aligned}
\end{equation}

We assumed the density value for all objects $\rho =$ 3000 $kg \cdot m^{-3}$ (as typical of V type asteroids \cite{Carruba2014}, \cite{moskovitz2008distribution}) and the Yarkovsky efficiency parameter $f_Y=10^{-5} \pm 0.5\cdot10^{-5} $ \citep{Nugent2012}. The drift rates are presented in table \ref{Yarko} and are only a first step approximation. More precise results can be obtained by long-term numerical integrations with the set of well determined thermal and rotational parameters of each asteroid. Most of our targets have maximum drift values on the order of $10^{-4}$  AU/My. We note the the age of Vesta collisional family is estimated as 1-2 Gy (\cite{spoto2015asteroid} \cite{carruba2016constraints}).
The age of the two large craters (Rheasilvia, Veneneia) is estimated as 1 Gy and 2 Gy respectively (e.g., \cite{marchi2012violent} \cite{schenk2012geologically}). This would mean maximum total Yarkovsky drift on the order of 0.1 AU for our targets during the Vesta family lifetime.

 \begin{table}[h]
\begin{center}
\begin{tabular}{|l|c|c|}
\hline
Asteroid     &   $\frac{da}{dt} \times 10^{-4}$ & $\Delta\frac{da}{dt}\times 10^{-4}$   \\
             &   $[AU/My]$ & $[AU/My]$   \\
\hline
(222) Lucia                     &      0.2              &      0.1 \\ 
(2704) Julian Loewe     &    -1.7                       &      0.9 \\
(4796) Lewis                    &      2.6              &      1.3  \\
(5150) Fellini                  &      2.3              &      0.7 \\
(5525) 1991 TS4                 &     $\pm$ 2.1         &      0.6 \\
(5599) 1999 SG1                 &     1.9                       &      0.9 \\
(5754) 1992 FR2                 &      1.9              &      0.5 \\
(5875) Kuga                     &      -1.6             &      0.4 \\
(6406) (1992 MJ)                &      -2.4             &      1.2 \\
(18641) 1998 EG10       &      -3.2             &      1.3 \\
\hline                                                  
\end{tabular}
\caption{Estimated maximum possible Yarkovsky drift.\label{yarko-drift}}
\label{Yarko}
\end{center}
\end{table}

\subsection{Estimating YORP effect}

Asteroid spin can be modified though the so-called Yarkovsky-O'Keefe-Radzievskii-Paddack effect (YORP). We have estimated the YORP rotational acceleration for our targets following \cite{Rozitis2013} and \cite{Rossi2009}:

\begin{equation}
\frac{d\omega}{dt}=\frac{G_1}{a^2 \sqrt{1-e^2} \rho D^2}C_Y
\end{equation}

which are shown in Table \ref{yorp-rate}. We assumed the max/min values of the YORP coefficient $-0.025 \le C_Y \le 0.025$ \citep{Rozitis2013}. 
Parameter $G_1$ is a modified solar constant ($-6.4 \times 10^{10}$ kg km s$^{-2}$), $a$ is the asteroid orbital semimajor axis in km, $e$ is the orbital eccentricity, $\rho$ is the bulk density in kg m$^{-3}$, $D$ is the diameter in m.

\begin{table}[h]                                                                             
\begin{center}                                                                               
\begin{tabular}{|l|c|c|}                                                                     
\hline                                                                                       
Asteroid     &   $\frac{d\omega}{dt} \times 10^{-19}$ & $\Delta\frac{d\omega}{dt}  \times 10^{-19}$   \\      
                     &  $[rad/{s^2}]$ & $[rad/{s^2}]$   \\      
\hline                                                                                       
(222) Lucia                     & $\pm$ 0.01 &  0.001   \\                                              
(2704) Julian Loewe     & $\pm$ 0.8  &  0.6    \\                                
(4796) Lewis                    & $\pm$ 2.0  &  1.3    \\                                   
(5150) Fellini                  & $\pm$ 1.3  &  0.2    \\                                   
(5525) 1991 TS4                 & $\pm$ 1.4  &  0.1    \\                              
(5599) 1999 SG1                 & $\pm$ 1.0  &  0.7    \\                              
(5754) 1992 FR2                 & $\pm$ 1.1  &  0.1    \\                                   
(5875) Kuga                     & $\pm$ 0.8  &  0.1    \\                                   
(6406) (1992 MJ)                & $\pm$ 1.8  &  1.2    \\                                   
(18641) 1998 EG10       & $\pm$ 3.1  &  1.3    \\                                   
\hline                                                                                       
\end{tabular}                                                                                
\caption{Estimated maximum YORP rates.}
\label{yorp-rate}     
\end{center}                                                                                 
\end{table}

We also estimated the YORP timescale (the time it takes for the asteroid rotation rate to be
doubled or halved) in Table \ref{yorp-time} (\citep{Rozitis2013}):

\begin{equation}
t_{YORP}=\frac{\omega}{|\frac{d\omega}{dt}|}.
\end{equation}

\begin{table}[h]                                                                             
\caption{Estimated YORP timescale.\label{yorp-time}}                     
\begin{center}                                                                               
\begin{tabular}{|l|c|}                                                                     
\hline                                                                                       
Asteroid  & $t_{YORP} \times 10^{7}$ [y]   \\      
\hline                                                                                       
(222) Lucia                     & 929.2  \\                                      
(2704) Julian Loewe     & 31.4  \\                                
(4796) Lewis                    & 7.9  \\                                   
(5150) Fellini                  & 8.0  \\                                   
(5525) 1991 TS4                 & 2.8  \\                              
(5599) 1999 SG1                 & 15.5  \\                              
(5754) 1992 FR2                 & 5.7  \\                                   
(5875) Kuga                     & 13.1  \\                                   
(6406) (1992 MJ)                & 4.5  \\                                   
(18641) 1998 EG10       & 3.4  \\                                                          
\hline                                                                                       
\end{tabular}                                                                                
\end{center}                                                                                 
\end{table}

We note that the calculated maximum YORP rates/YORP times are only approximate statistical values. The YORP timescales for our targets are on the order of around $\sim$0.01$\times$A$_{Vesta}$ (where A$_{Vesta}$ is Vesta family age).

If we assume maximum values of YORP torques, referred to the rate change in obliquity $\frac{d \epsilon}{dt}$ (also marked in literature as  $\frac{d \xi}{dt}$) 
we can calculate the maximum possible obliquity shift.
According to \cite{Bottke2006}, the rates of change of $\omega$, $\epsilon$ and $\psi$ can be calculated by
\begin{equation}
\frac{d\omega}{dt}=\frac{\mathbf{T} \cdot \mathbf{e}}{C}=\frac{T_{S}}{C}
\end{equation} 
\begin{equation}
\frac{d\epsilon}{dt}=\frac{\mathbf{T} \cdot \mathbf{e}_{\perp 1}}{C\omega}=\frac{T_{\epsilon}}{C\omega}
\end{equation} 
\begin{equation}
\frac{d\psi}{dt}=\frac{\mathbf{T} \cdot \mathbf{e}_{\perp 2}}{C\omega}=\frac{T_{\psi}}{C\omega},
\end{equation} 
where $T_{S}$, $T_{\epsilon}$, $T_{\psi}$ are YORP torque components, 
$\mathbf{T}$-torque , $\mathbf{e}$ - spin axis unit vector, decomposed into orbital plane unit vectors $\mathbf{e}_{\perp 1}$, $\mathbf{e}_{\perp 2}$  \citep{Bottke2006}.
We can `maximize' obliquity drift by assuming that the $T_{\epsilon}$ component has the maximum possible value ($\mathbf{e} \simeq \mathbf{e}_{\perp 1}$). 
Assuming maximum possible values of $T_{\epsilon}$ \citep{Bottke2006}, we obtain the values listed in Table \ref{obliqrate}.

\begin{table}[h]                                                                             
\caption{Estimated maximum obliquity rates.}                     
\label{obliqrate}
\begin{center}                                                                               
\begin{tabular}{|l|c|c|}                                                                     
\hline                                                                                       
Ast.   name,     &   $\frac{d\epsilon}{dt} \; [deg/{My}]$ & $\Delta\frac{d\epsilon}{dt} \; [deg/{My}]$   \\      
\hline                                                                                       
(222) Lucia                     & 0.01  & 0.001     \\                                      
(2704) Julian Loewe     & 0.18  & 0.12     \\                                
(4796) Lewis                    & 0.73  & 0.49     \\                                   
(5150) Fellini                  & 0.72  & 0.11     \\                                   
(5525) 1991 TS4                 & 2.04  & 0.22     \\                              
(5599) 1999 SG1                 & 0.37  & 0.25     \\                              
(5754) 1992 FR2                 & 1.00  & 0.09     \\                                   
(5875) Kuga                     & 0.44  & 0.05     \\                                   
(6406) (1992 MJ)                & 1.26  & 0.85     \\                                   
(18641) 1998 EG10       & 1.71  & 0.71     \\                                                      
\hline                                                                                       
\end{tabular}                                                                                
\end{center}                                                                                 
\end{table}

The estimated maximum obliquity changes for most of our targets are less than 1 deg per My. The estimated upper limit for the YORP rates, obliquity rates and lower limit on the YORP time-scale do not exclude spin reorientation (including change in the direction of rotation) due to YORP. More detailed studies are however necessairy to obtain more realistic estimates for those effects. In particular further observations and determining the detailed spin and shape properties might help in constraining the magnitude of those effects.

\subsection{Collisional reorientation timescales}
Asteroid spins can be modified by random collisions. We estimate the collisional reorientation timescales for our targets to assess the importance of collisions in our sample. Assuming presence of nondisruptive collisions between asteroids, we can estimate the characteristic timescale of these events (time of spin axis reorientation), from the nominal formula of \cite{farinella1998meteorite}: 
\begin{equation}
\label{eq1}
\tau_{reor} \simeq BR^{\beta},
\end{equation}
where the coefficients typicaly are $B \simeq 15$ Myr and $\beta=0.5$ (\cite{Morbidelli2003}, \cite{Bottke2006}).  However because radius $R$ and frequency $\omega$ are estimated with high precision for our targets, 
we can use more complicated, $\omega$-dependent reorientation formula \citep{Morbidelli2003}:

\begin{equation}
T_{reor}=15\left(\frac{1}{500}\frac{\omega}{\omega_{0}}\right)^{\frac{5}{6}}R^{\frac{4}{3}} \; Myr,
\end{equation}

where $\omega_{0}=2\pi/(5h)$. This formula can be reduced to nominal reorientation law mentioned above (Eq. \ref{eq1}) if we assume that $\omega$ scales as $1/R$ and $P=5$ h for $R=500$ m \citep{Morbidelli2003}.

\begin{table}[h]                                                                             
\caption{Estimated reorientation timescales.\label{reor-time}}                     
\begin{center}                                                                               
\begin{tabular}{|l|c|}                                                                     
\hline                                                                                       
Ast.   name,  & $T_{reor}$ [$\times 10^7$ y]   \\      
\hline                                                                                       
(222) Lucia          & 5020  \\                                   
(2704) Julian Loewe  & 943    \\                                
(4796) Lewis         & 352   \\                                   
(5150) Fellini       & 308  \\                                   
(5525) 1991 TS4      & 151  \\                              
(5599) 1999 SG1      & 528  \\                              
(5754) 1992 FR2      & 256  \\                                   
(5875) Kuga          & 449  \\                                   
(6406) (1992 MJ)     & 229  \\                                   
(18641) 1998 EG10    & 185  \\                                   
\hline                                                                                       
\end{tabular}                                                                                
\end{center}                   
\label{reorient}                                                              
\end{table}

The estimated reorientation timescales are listed in Table \ref{reorient}. The estimated reorientation timescales for all objects are longer than $10^{9}$ yr, that is longer then the estimated age of the Vesta family. 
Therefore statistically collisions are unlikely to modify spins of substantial part of the population.

\section{Conclusions}
\label{concl}

\begin{figure}[htbp]
   \centering
   \includegraphics[width=1.0\linewidth]{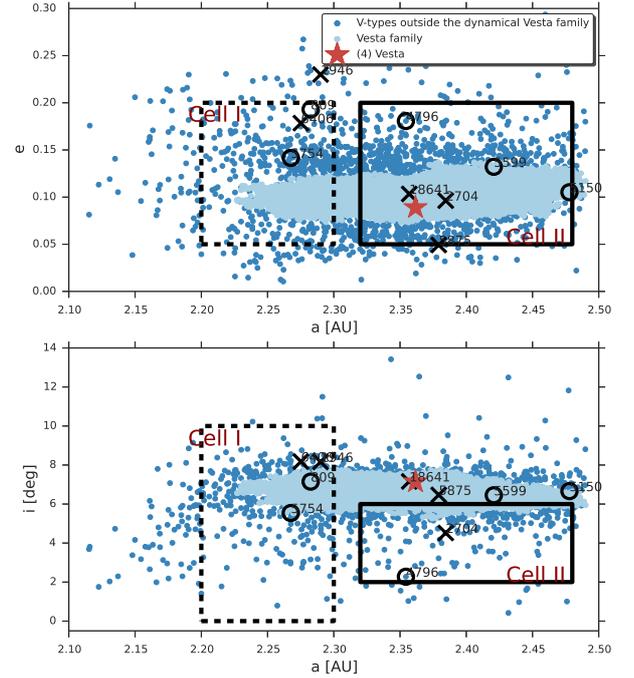} 
   \caption{Distribution of the prograde (circles) and retrograde (crosses) V-type rotators in the inner main asteroid belt. Location of asteroid (4) Vesta is denoted with a red star. Vesta family and V-type candidates outside the Vesta family are marked. Cell I (dashed-line square) and Cell II (solid-line square) regions are denoted as in \citep{nesvorny2008fugitives}. }
   \label{comb}
\end{figure}

We have determined sense of rotation for eight asteroids (including the test object (222) Lucia). For three objects the spin orientation was known from the literature. For one object ((5525) 1991 TS4) we were not able to determine its sense or rotation. Overall our sample included eleven V-type asteroids outside the dynamical Vesta family. The distribution of the prograde and retrograde rotators in orbital element space is shown in Fig. \ref{comb}. 

Assuming that the direction of rotation of our asteroids was not reversed due to YORP effect we can speculate on their origin.
Five of the observed objects ((1946) Walraven, (5150) Fellini, (5599) 1991TS4, (6406) 1992 MJ and (18641) 1998EG10 can be directly explained by migration from the Vesta family.  
The origin of three other objects ((2704) Julian Loewe, (4796) Lewis, (5875) Kuga) is ambiguous and dynamical studies are needed to better understand their origin. Two asteroids ((809) Lundia and (5754) 1992 FR2) are the least likely to have a genetic connection to Vesta. Those two objects are located in Cell I, have large diameters ($\sim$ 9.1 km and $\sim$ 6.5 km respectively) and have prograde rotation (positive Yarkovsky drift) therefore must have migrated to their current positions from the direction opposite to Vesta. 

The estimated collisional lifetimes of the objects are on the order of 10$^{9}$ yr and longer then the age of the Vesta family. Thus we conclude that random collisions are unlikely to modify spins of substantial part of the population. 
The estimated upper limit for the YORP rates and obliquity rates does not exclude spin modification due to YORP. Further studies should be conducted to determine the detailed spin properties of those objects and estimate more realistic YORP rates.
In particular, the candidate non-Vestoids should be primary targets for further observations and more realistic YORP estimates.

The distribution of the five misfitting objects (ambiguous and non-Vestoid candidates) is scattered, similar to the V-type distribution in the mid and outer main belt. If a non-link to Vesta can be confirmed, this distribution would qualitatively fit well with the \cite{bottke2006iron} theory. To test the theory a larger number of inner belt V-types has to be surveyed with respect to sense or rotation. Prograde rotators in Cell I are particularly strong non-Vestoid candidates. If the \cite{bottke2006iron} theory is valid, there should be more non-Vestoids scattered in the inner main belt than in the mid and outer main belt. Therefore determining the ratio of the prograde rotators in Cell I to non-Vestoids in the mid and outer main belt is the ultimate goal of the research started here.

Similarly the low inclination V-types were not reproduced by \cite{nesvorny2008fugitives}. Determining the ratio of prograde to retrograde rotators in this population and comparing it to the ratio produced by numerical migration simulations and to the spin-state ratio in the Vesta family may help understand the origin of this population.

\section{Acknowledgments}
We would like to acknowledge support from Lowell Observatory. This article is based on 109 nights of observations made with the 1.1m Hall, 1.8m Perkins and 0.8m telescopes at Lowell Observatory. We would like to express our gratitude to Lowell Observatory staff (especially to Len Bright, Larry Wasserman, Peter Collins, and Ted Dunham) for technical support during the observing nights. The remaining observations (three nights) were made with the IAC-80 telescope operated on the island of Tenerife by the Instituto de Astrof\'{\i}sica de Canarias in the Spanish Observatorio del Teide. J. Licandro acknowledge support from the AYA2015-67772-R  (MINECO, Spain). The work of A. Marciniak was supported by grant no. 2014/13/D/ST9/01818 from the National Science Centre, Poland.

\begin{landscape}
\begin{center}
\begin{table}[htbp]
   \begin{tabular}{|l l l l l ll  l  l ll ll | }  \hline 
   Asteroid name                & H                     & a             & e               & i                     & Opp. date             & d                                     & $\sim$ $\Delta$v                        & P$_{Ref}$                                     & P$_{Oszk}$ (h)                  & Rot. sense            & Yark. drift                                            &  \\ 
   (Cell location)              & (mag)         & (AU)  &               & ($^{\circ}$)    &                               & (km)                          & (m/s)                                   & (h)                                           &                                               &                               &                                                               &  \\ \hline       
   (222) Lucia                  & 9.13          & 3.14  & 0.13  & 2.14          & 2016/04/14              & 56.52$\pm 0.83$       & -                                             & 7.8373 $\pm$0.0003 (1)          & 7.8371 $\pm$ 0.0002           & Prograde              & NA                                                              & \\
   (809) Lundia (I)             & 12.2          & 2.28  & 0.19  & 7.15          & -                               & 9.1                                   & 3586                                    & 15.4142$\pm$0.0006 (1,2)              & -                                               & Prograde              & Toward Vesta                                    & \\
   (1946) Walraven (I)  & 12.0          & 2.29  & 0.23  & 8.16          & -                               & 10.0                          & 2659                                  & 10.2101$\pm$0.0005 (3)          & -                                             & Retrograde              & Away from Vesta                                       & \\
   (2704) Julian Loewe (II)     & 12.7          & 2.38  & 0.10  & 4.52          & 2016/08/09              & *7.08$\pm 2.14$               &  813                                  & 2.091$\pm$0.001 (4)                     & 2.6383$\pm$ 0.0001            & Retrograde              & Toward Vesta                                  & \\
   (4796) Lewis (II)            & 13.6          & 2.35  & 0.18  & 2.27          & 2016/03/17              & *4.68$\pm1.41$                &  1949                                 & 3.5080$\pm$0.0002 (5,6)         & 3.5086$\pm$0.0001             & Prograde              & Toward Vesta                                    & \\
   (5150) Fellini                       & 13.3          & 2.48  & 0.11  & 6.66            & 2016/03/11            & 5.401$\pm 0.229$              &  1462                           & 5.8270$\pm$0.0008 (7)                 & 5.1953$\pm$0.0002               & Prograde              & Away from Vesta                                       & \\
   (5525) 1991 TS4 (I)  & 13.2          & 2.22  & 0.15  & 7.6                   & 2016/04/19              & 5.917$\pm 0.115$              &  1558                                 & 6.972$\pm$0.003 (1)                     &14.088$\pm$0.004                       & -                               & -                                                             &  \\     
   (5599) 1999 SG1              & 12.9          & 2.42  & 0.16  & 6.44          & 2016/05/25              & *6.46$\pm1.95$                &  1690                                 & 3.62$\pm$0.01 (8)                               & 3.6194$\pm$0.0001             & Prograde                & Away from Vesta                                       & \\
   (5754) 1992 FR2 (I)  & 12.9          & 2.26  & 0.14  & 5.53          & 2016/04/17              & 6.577$\pm 0.064$              &  1188                         & 8.898$\pm$0.002 (1)                     & 8.90090$\pm$0.0005            & Prograde                & Toward        Vesta                                   & \\
   (5875) Kuga                  & 12.9          & 2.37  & 0.05  & 6.47          & 2016/01/26              &7.465$\pm 0.144$               &  964                                  & 5.551$\pm$0.002 (9)                     & 5.5512$\pm$0.0002             & Retrograde              & Toward Vesta                                  & \\
   (6406) (1992 MJ) (I) & 13.4          & 2.27  & 0.18  & 8.17          & -                               & *5.13$\pm1.55$                & 1328                                  & 6.81765$\pm$0.00001 (4)         & -                                             & Retrograde              & Away Vesta                                            &  \\
   (18641) 1998 EG10    & 14.0          & 2.35  & 0.10  & 7.16          & 2016/04/13              & 3.717$\pm 0.650$              & 656                                   & 5.68$\pm$0.01 (10)                      & 5.2461$\pm$0.0001             & Retrograde              & Away Vesta                                            &  \\ 
\hline
      \end{tabular}
   \caption{Absolute magnitude, orbital elements, opposition date, diameter (WISE or * estimated as $D = \frac{1329}{\sqrt{p}}10^{-0.2H}$), ejection velocity, literature rotational periods$^3$  and 
   periods estimated in this work, senses of rotation, and direction of the Yarkovsky drift for asteroids in this study.}
   \label{targets}
\end{table}
\end{center}

\footnote{References: (1) \url{http://obswww.unige.ch/~behrend/}, (2) \cite{kryszczynska2009new}, (3) \cite{hanuvs2015new}, (4) \cite{Clark2015}, (5) \cite{chang2014313},
   (6) \cite{hasegawa2014lightcurve}, (7) \cite{waszczak2015asteroid}, (8) \cite{2004Willis}, (9) \cite{2009Carbo}, (10) \cite{2008Clark}}
\end{landscape}

\twocolumn

\newpage
\bibliography{biblio}
\bibliographystyle{aa}

\newpage
\begin{appendix}
Below we present lightcurves of the observed asteroids and determination of the synodic periods.

\onecolumn
\section{Lightcurves}
\begin{figure}[ht]
  \centering
  \begin{subfigure}[b]{0.45\linewidth}
    \centering\resizebox{\hsize}{!}{\includegraphics[width=0.5\textwidth]{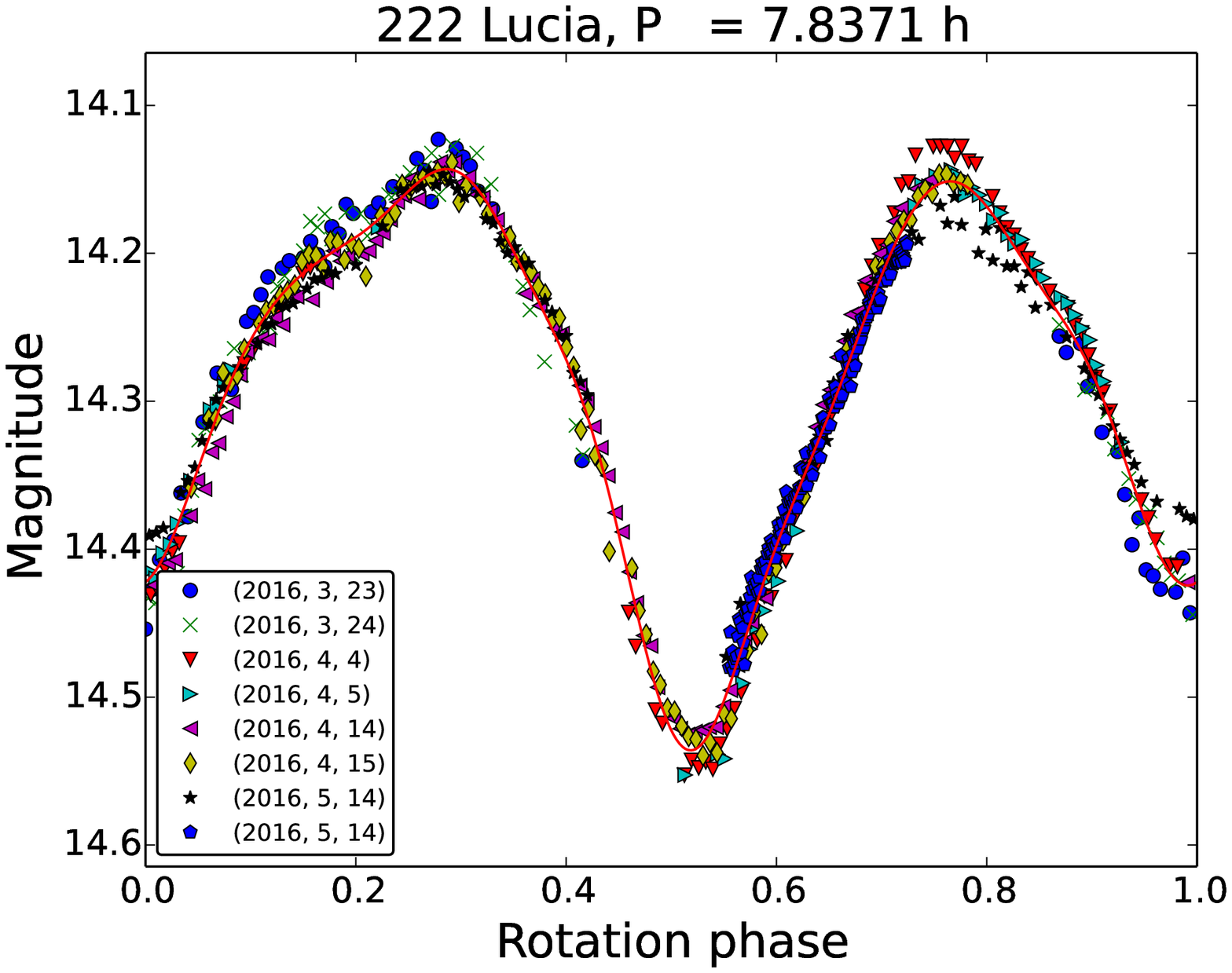}}
    \caption{Before opposition. \label{LuciaOpp}}
  \end{subfigure}  %
  \begin{subfigure}[b]{0.45\linewidth}
    \centering\resizebox{\hsize}{!}{\includegraphics[width=0.5\textwidth]{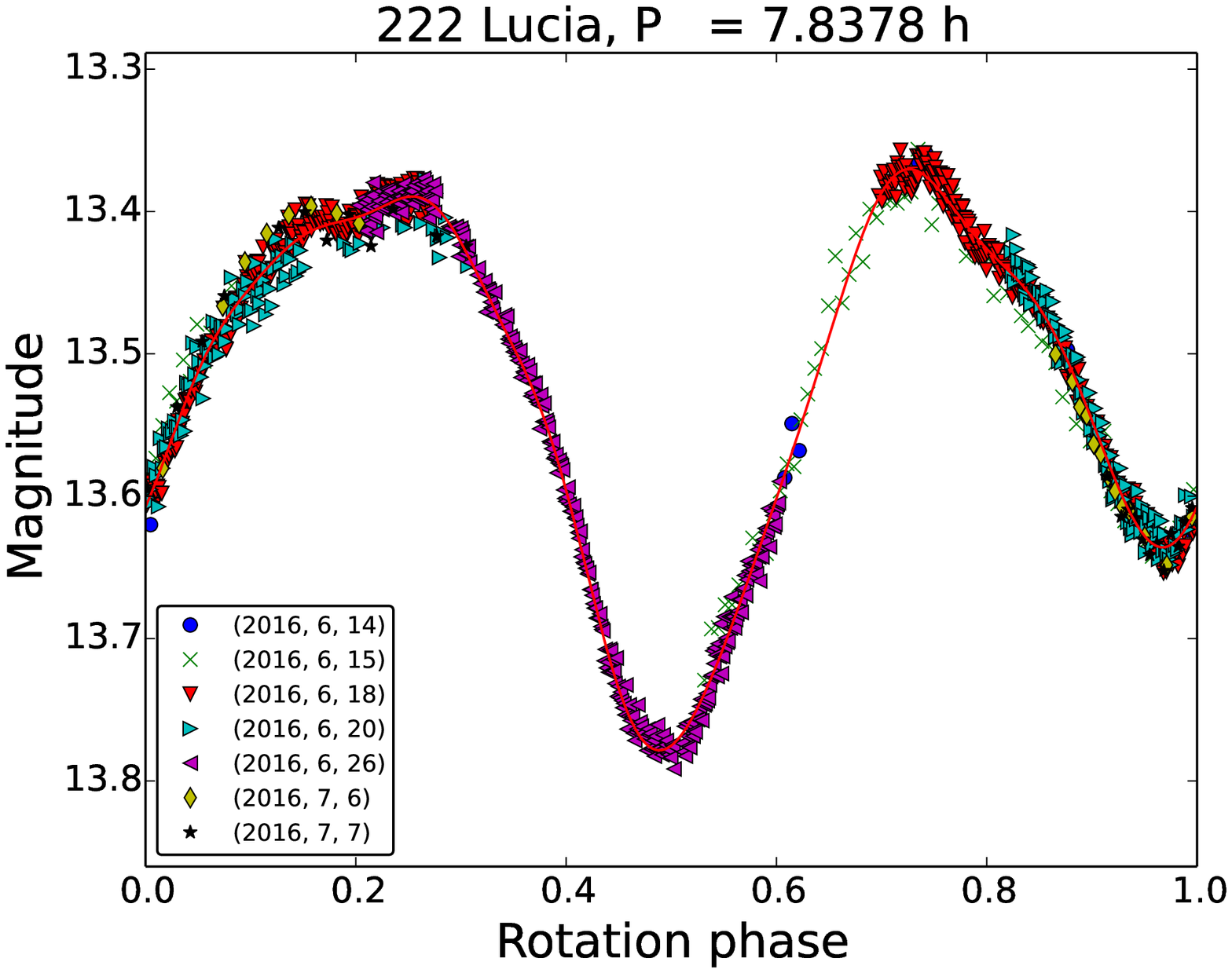}}
    \caption{At opposition. \label{LucisAfter}}
  \end{subfigure}
  \caption{Composite lightcurves for (222) Lucia.}
\end{figure}

\begin{figure}[ht]
  \centering
  \begin{subfigure}[b]{0.45\linewidth}
    \centering\resizebox{\hsize}{!}{\includegraphics[width=0.5\textwidth]{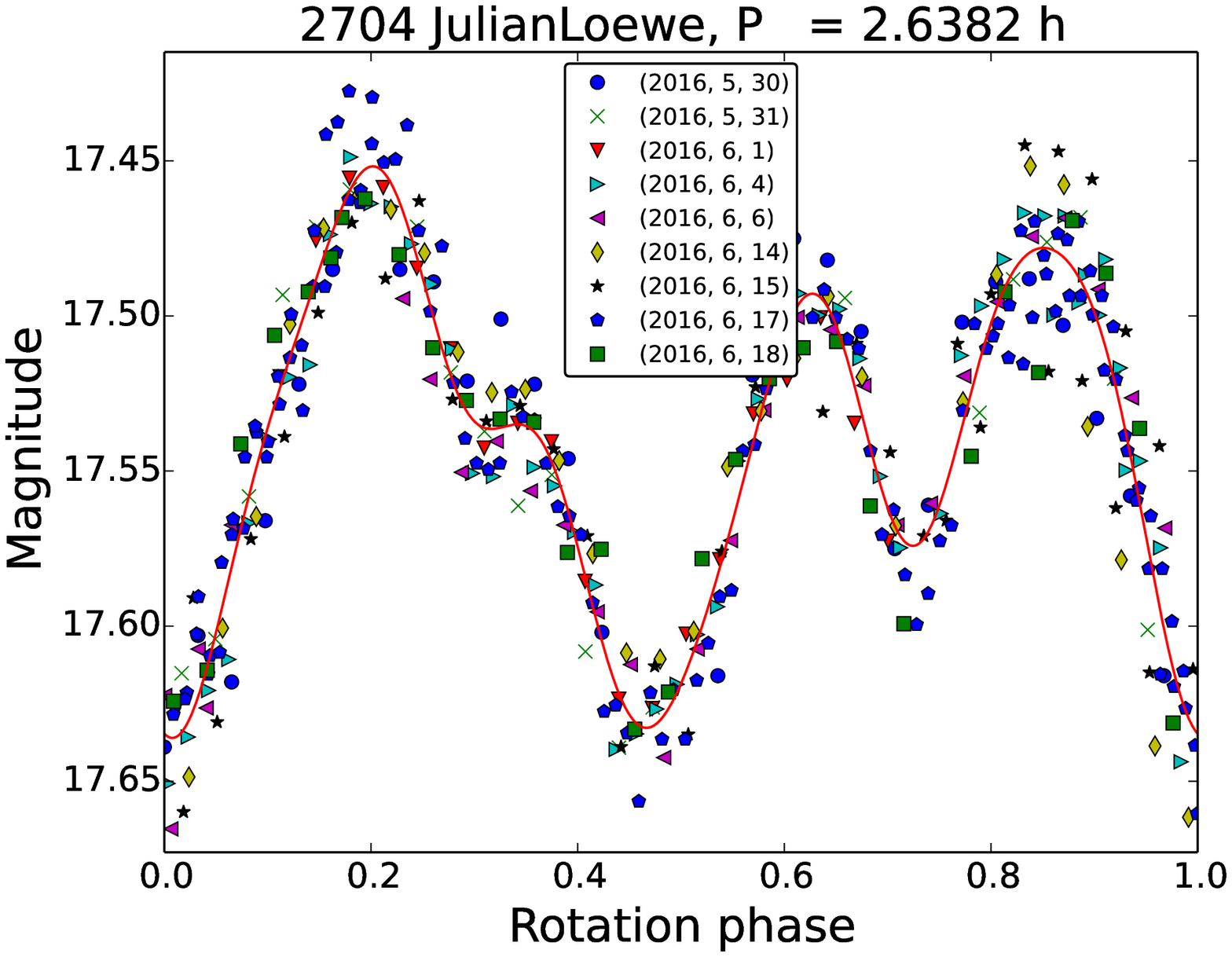}}
    \caption{Before opposition. \label{JulianOpp}}
  \end{subfigure}  
  \begin{subfigure}[b]{0.45\linewidth}
    \centering\resizebox{\hsize}{!}{\includegraphics[width=0.5\textwidth]{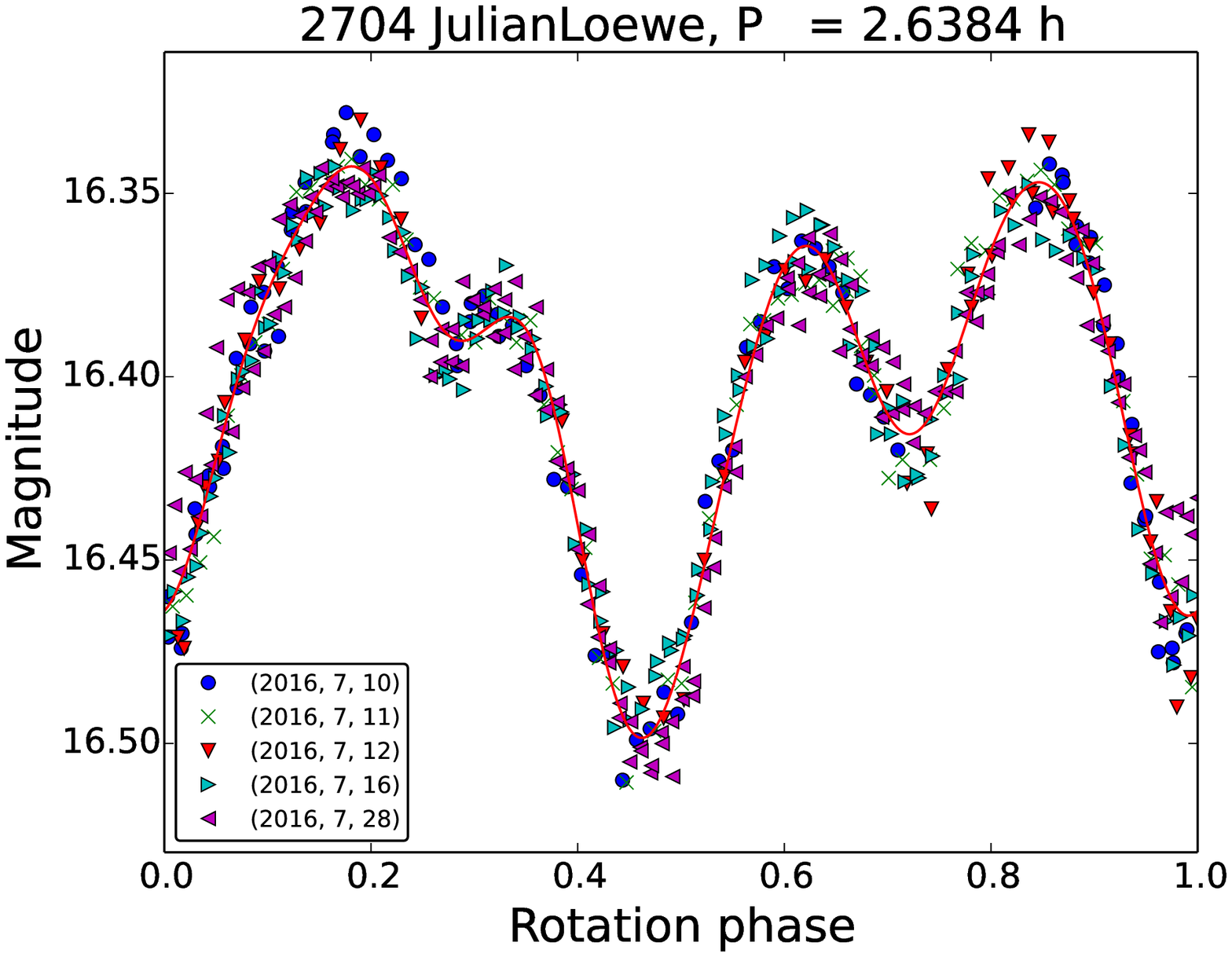}}
    \caption{At opposition. \label{JulianAfter}}
  \end{subfigure} \\
    \begin{subfigure}[b]{0.45\linewidth}
    \centering\resizebox{\hsize}{!}{\includegraphics[width=0.5\textwidth]{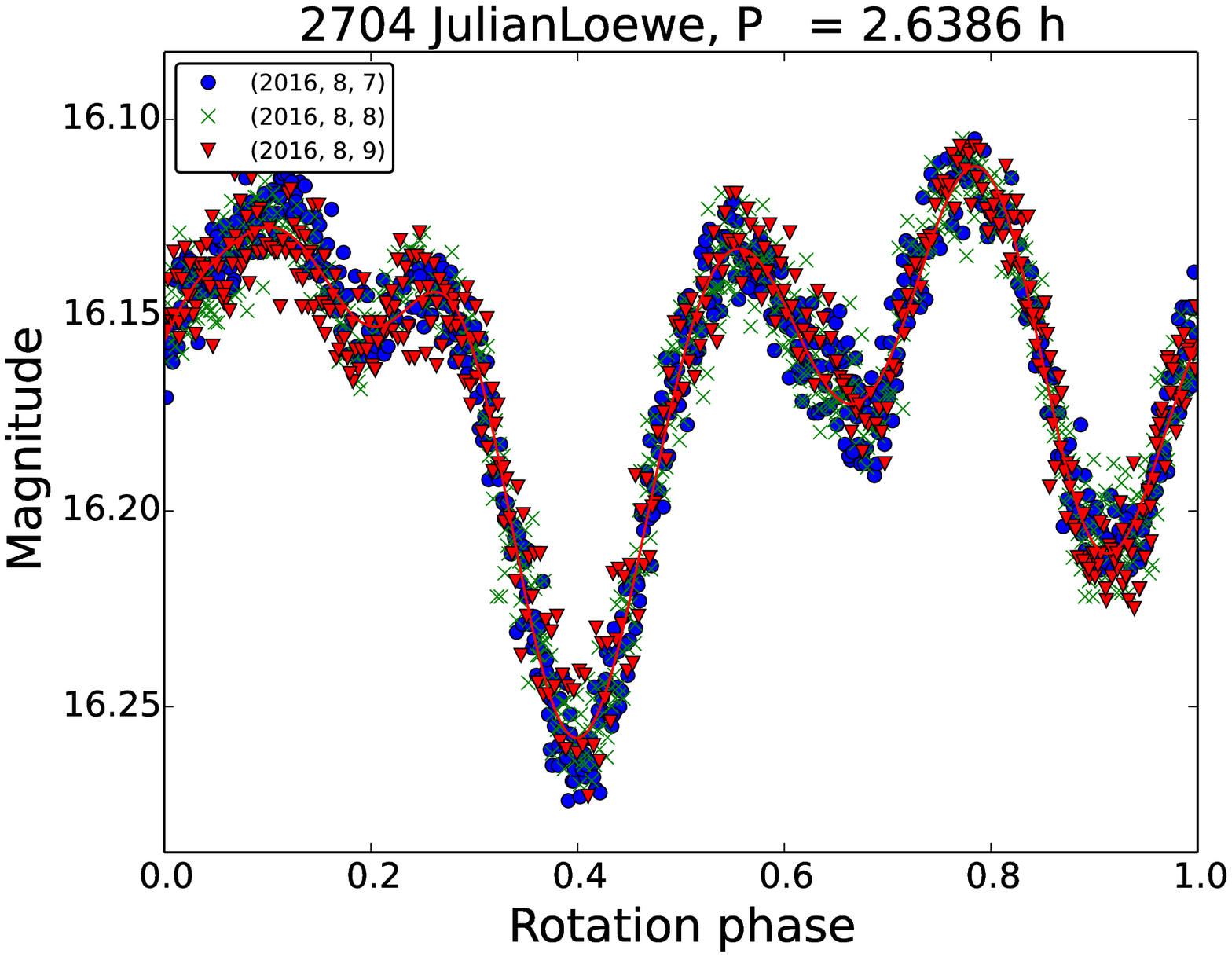}}
    \caption{At opposition. \label{JulianAfter}}
  \end{subfigure} 
     \begin{subfigure}[b]{0.45\linewidth}
    \centering\resizebox{\hsize}{!}{\includegraphics[width=0.5\textwidth]{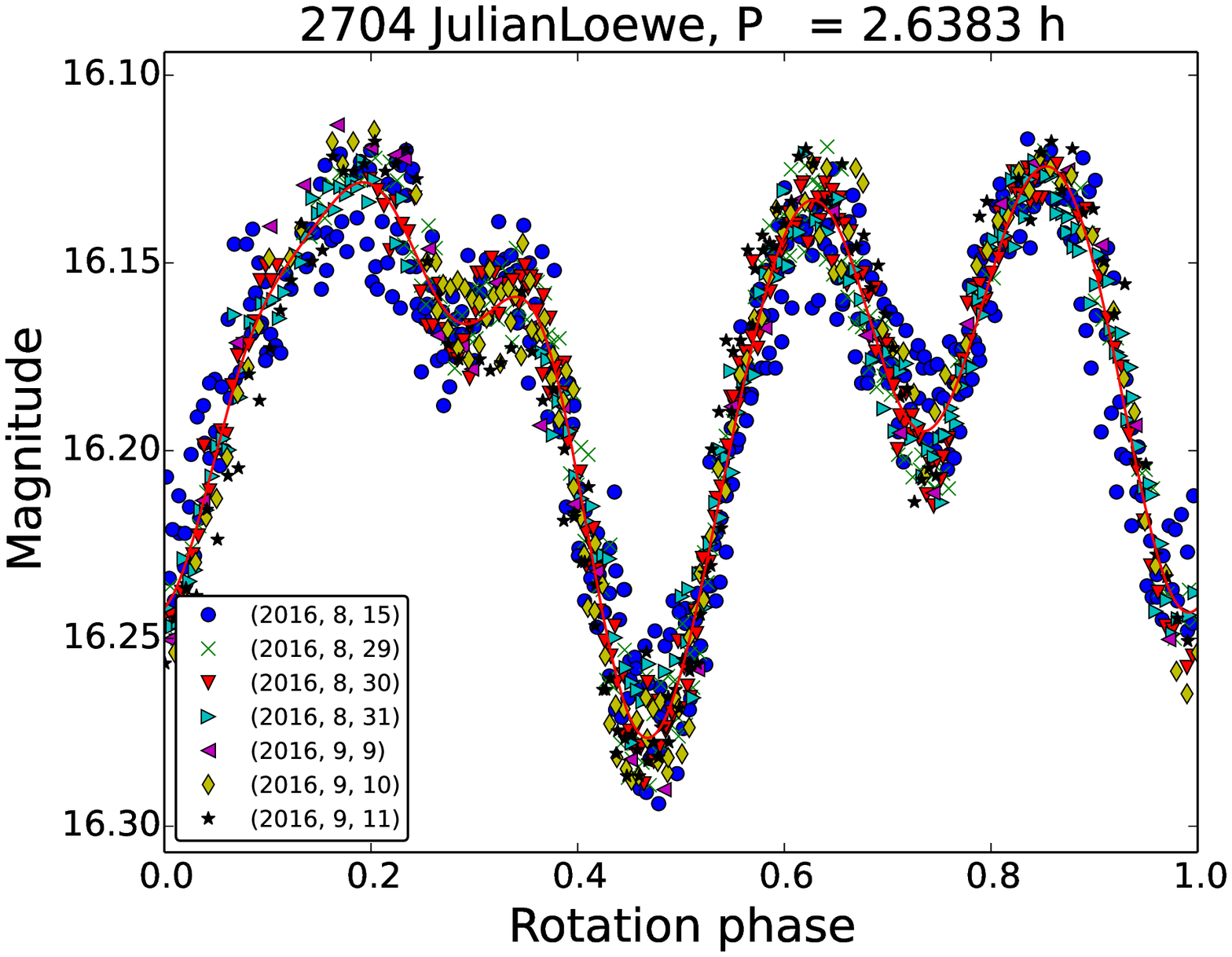}}
    \caption{After opposition. \label{JulianAfter}}
  \end{subfigure}
  \caption{Composite lightcurves for (2704) Julian Loewe. \label{loewe}}
\end{figure}

  \begin{figure}[ht]
  \centering
  \begin{subfigure}[b]{0.45\linewidth}
    \centering\resizebox{\hsize}{!}{\includegraphics[width=0.5\textwidth]{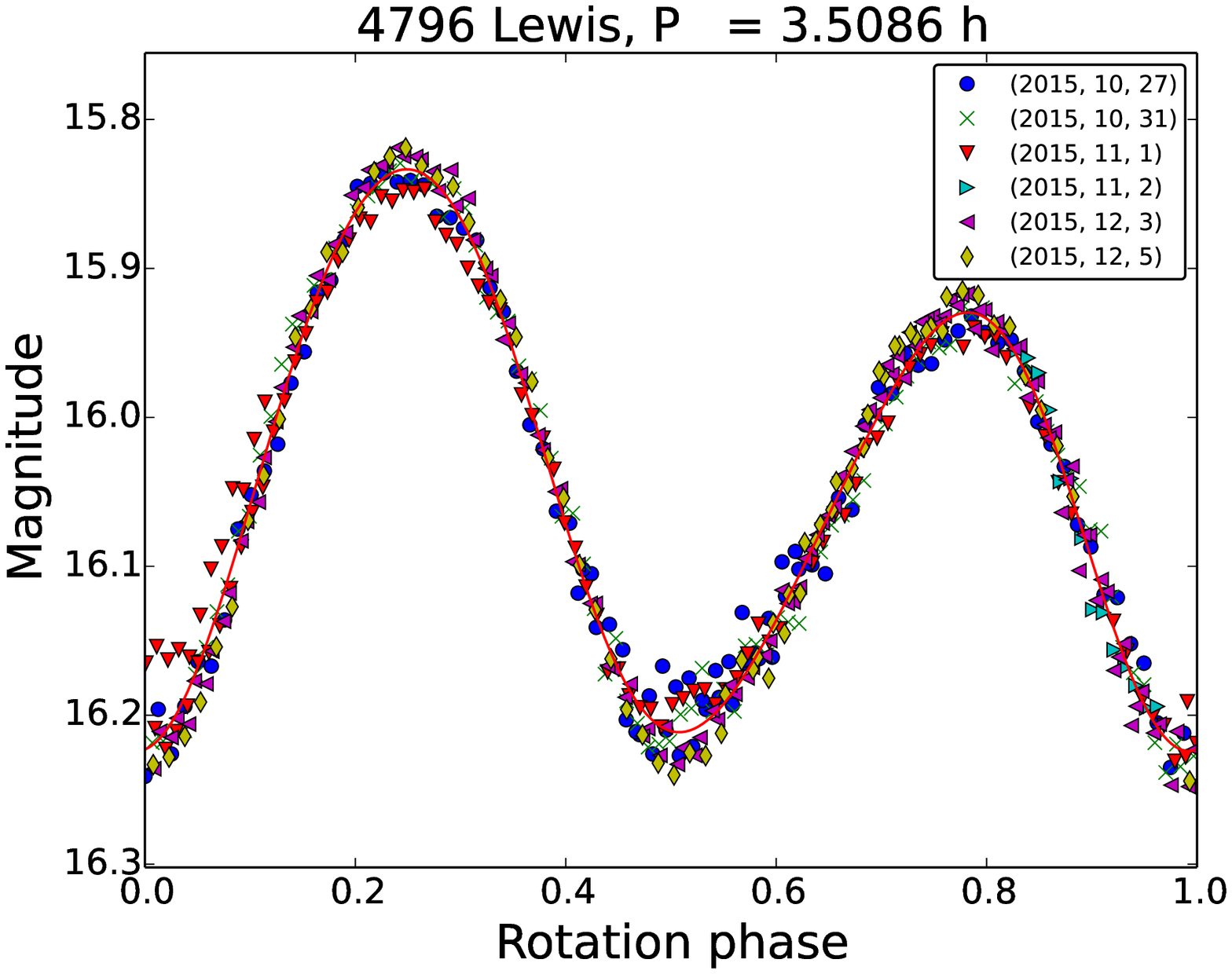}}
    \caption{At opposition. \label{LewisOpp}}
  \end{subfigure} %
  \begin{subfigure}[b]{0.45\linewidth}
    \centering\resizebox{\hsize}{!}{\includegraphics[width=0.5\textwidth]{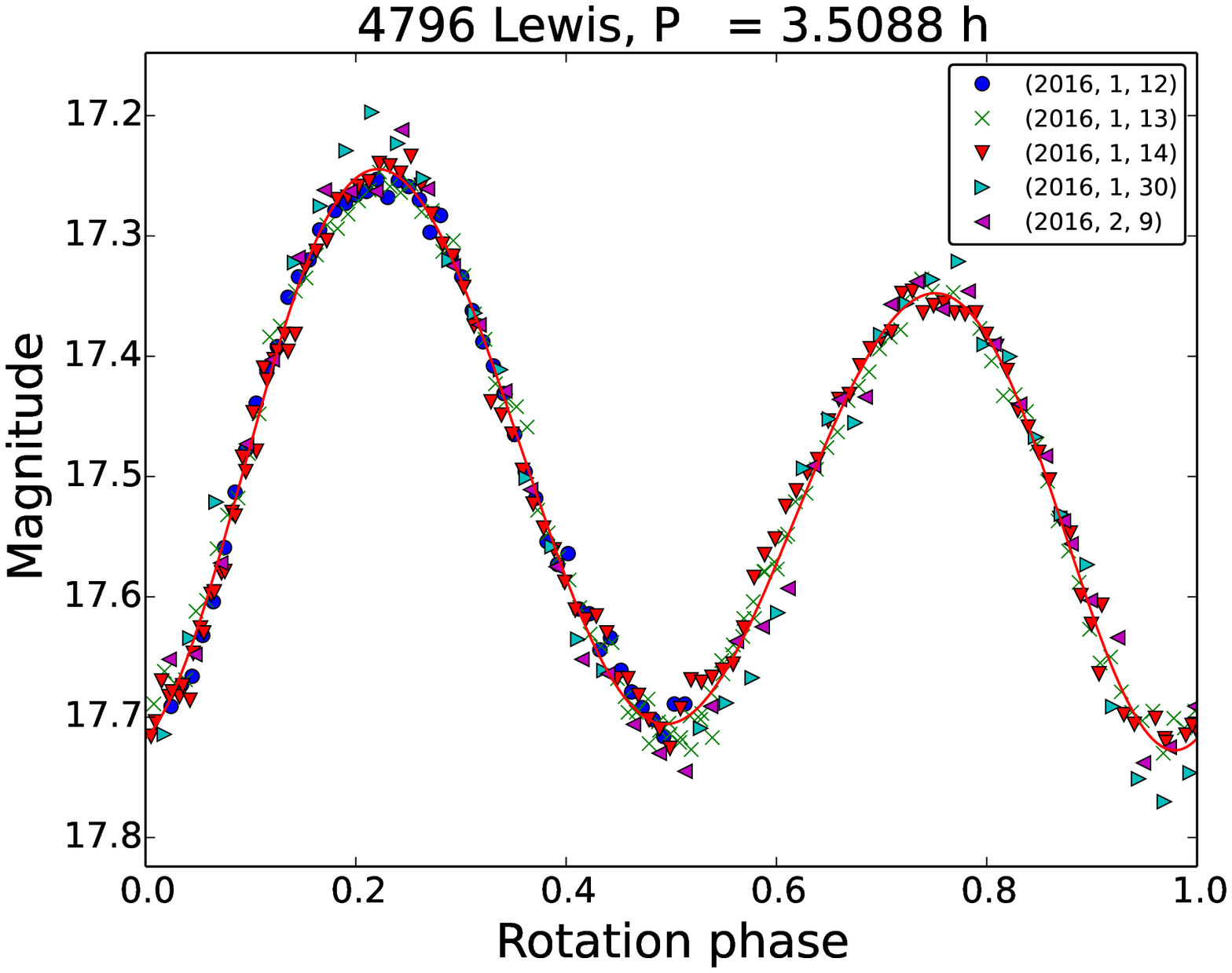}}
    \caption{After opposition. \label{LewisAfter}}
  \end{subfigure}
  \caption{Composite lightcurves for (4796) Lewis.}
\end{figure}

\begin{figure}[ht]
  \centering
  \begin{subfigure}[b]{0.45\linewidth}
    \centering\resizebox{\hsize}{!}{\includegraphics[width=0.5\textwidth]{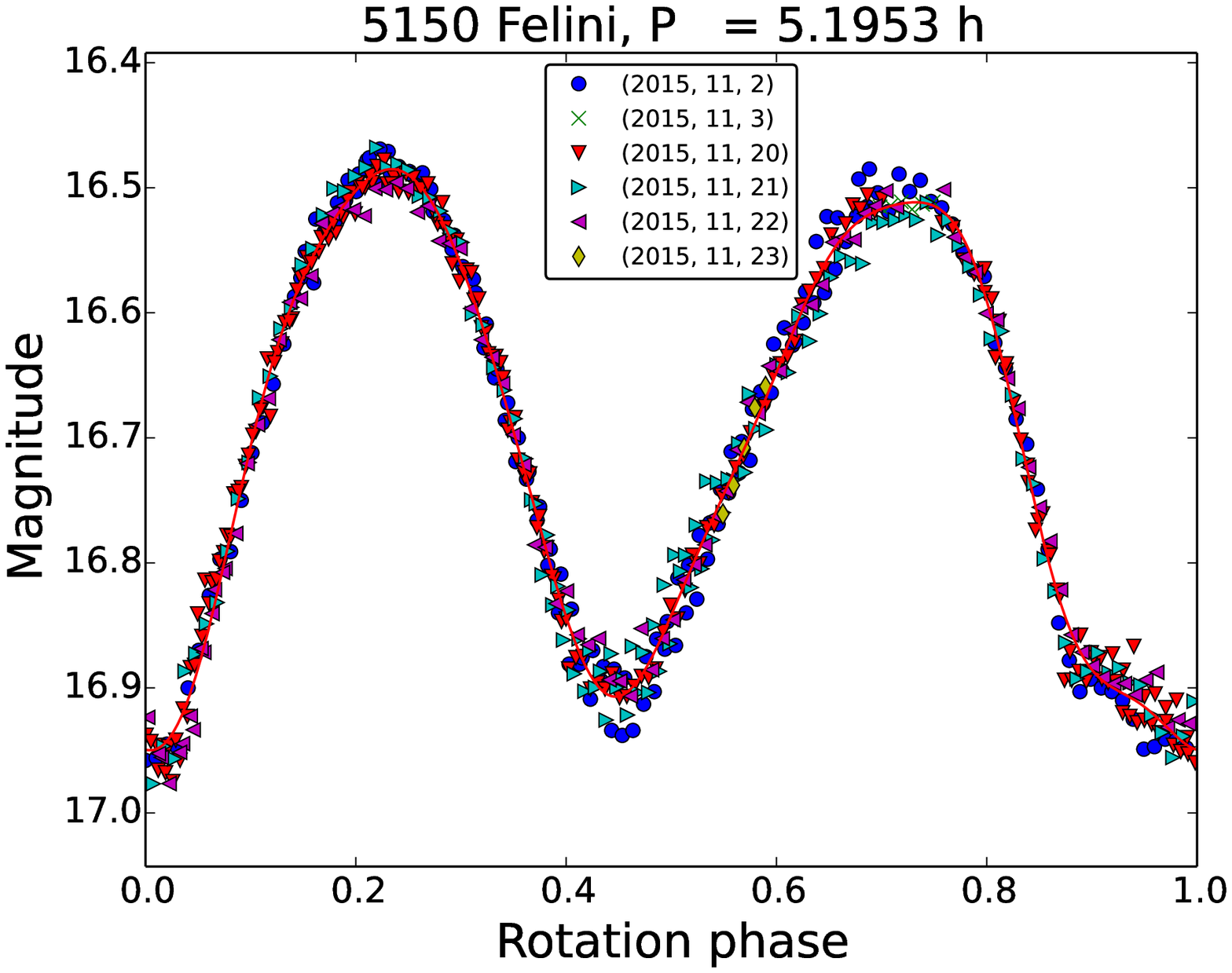}}
    \caption{At opposition. \label{FelliniOpp}}
  \end{subfigure} %
  \begin{subfigure}[b]{0.45\linewidth}
    \centering\resizebox{\hsize}{!}{\includegraphics[width=0.5\textwidth]{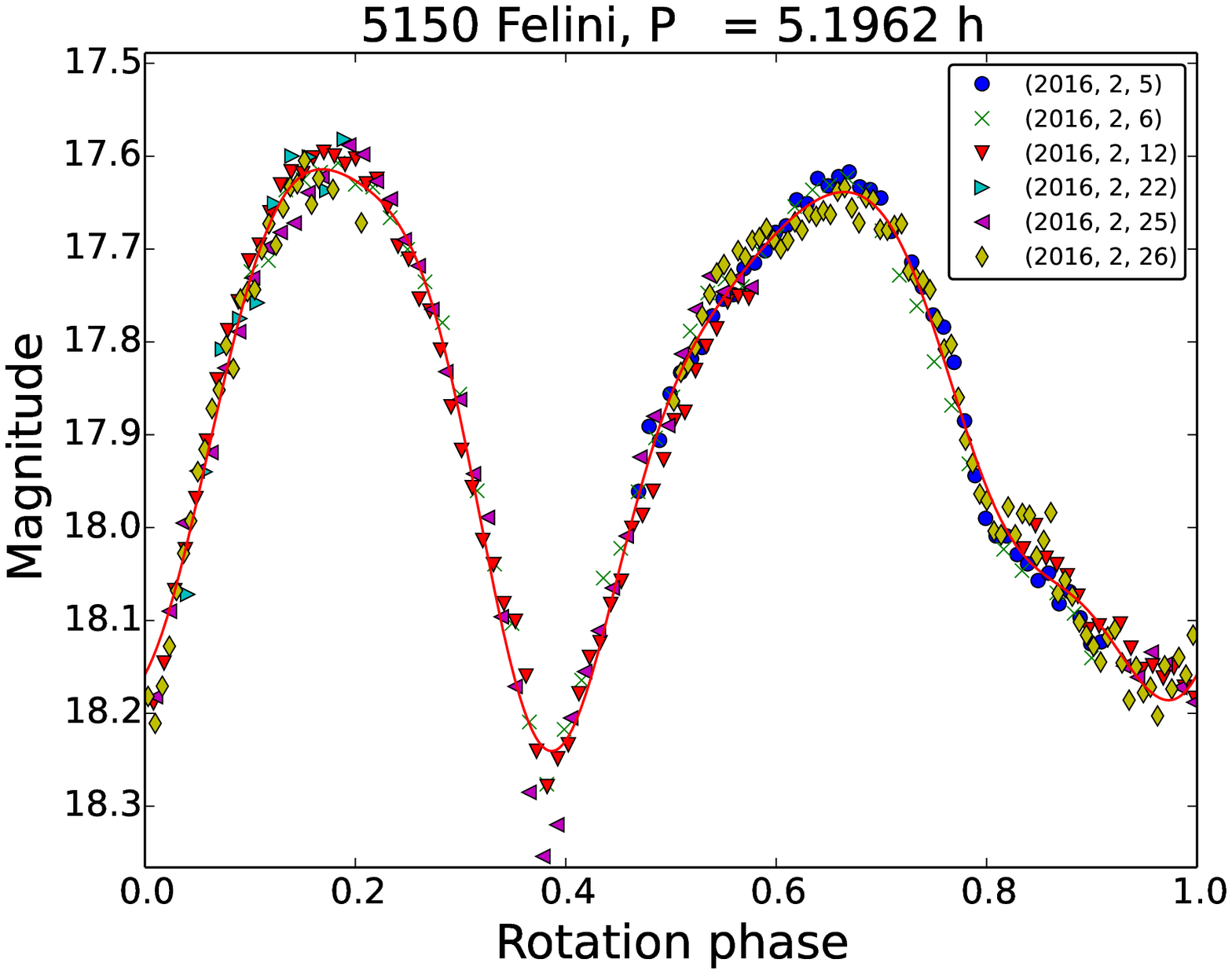}}
    \caption{After opposition. \label{FelliniAfter}}
  \end{subfigure}
  \caption{Composite lightcurves for (5150) Fellini.}
\end{figure}

\begin{figure}[ht]
  \centering
  \begin{subfigure}[b]{0.45\linewidth}
    \centering\resizebox{\hsize}{!}{\includegraphics[width=0.5\textwidth]{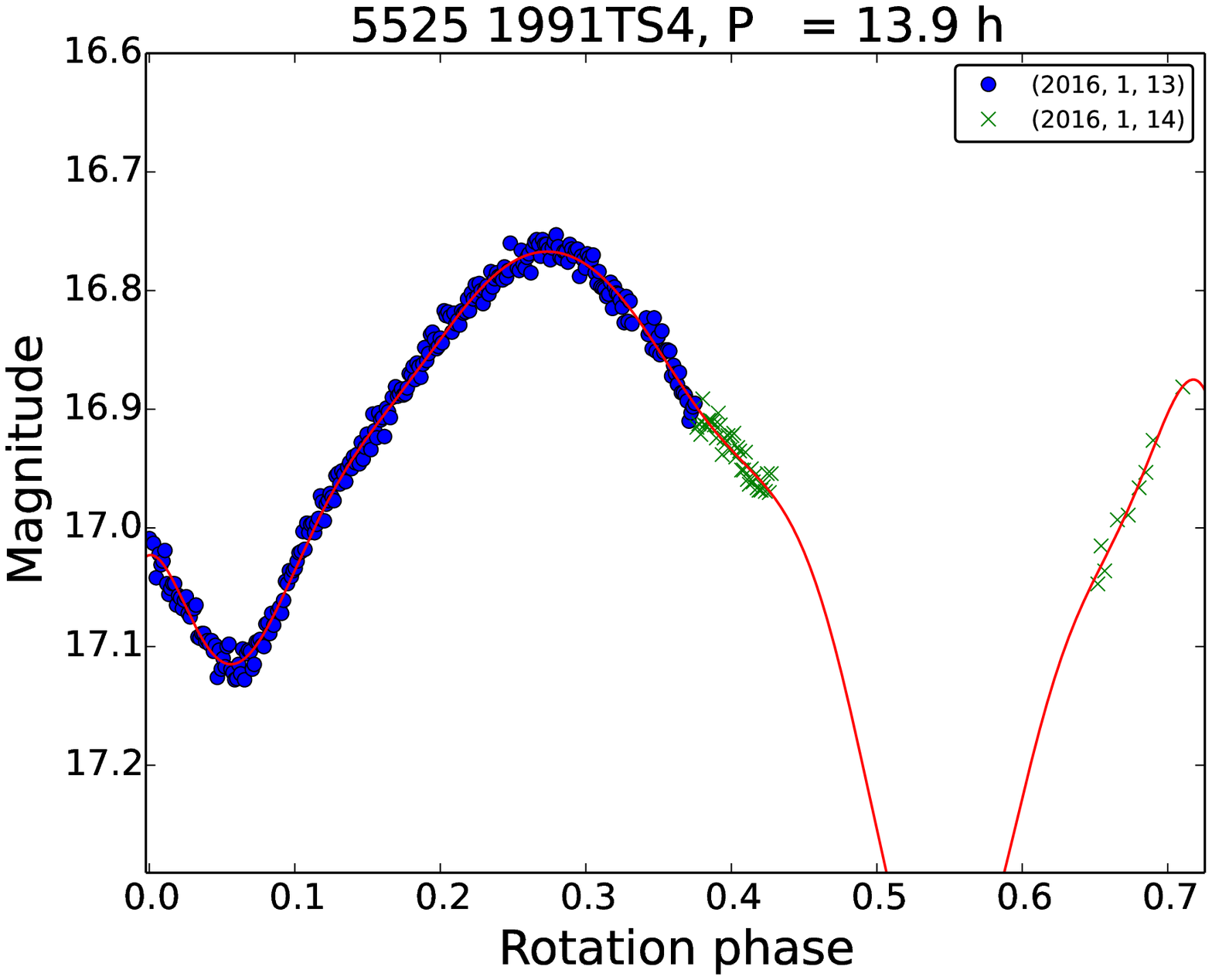}}
    \caption{At opposition. \label{5525b}}
  \end{subfigure}  %
  \begin{subfigure}[b]{0.45\linewidth}
    \centering\resizebox{\hsize}{!}{\includegraphics[width=0.5\textwidth]{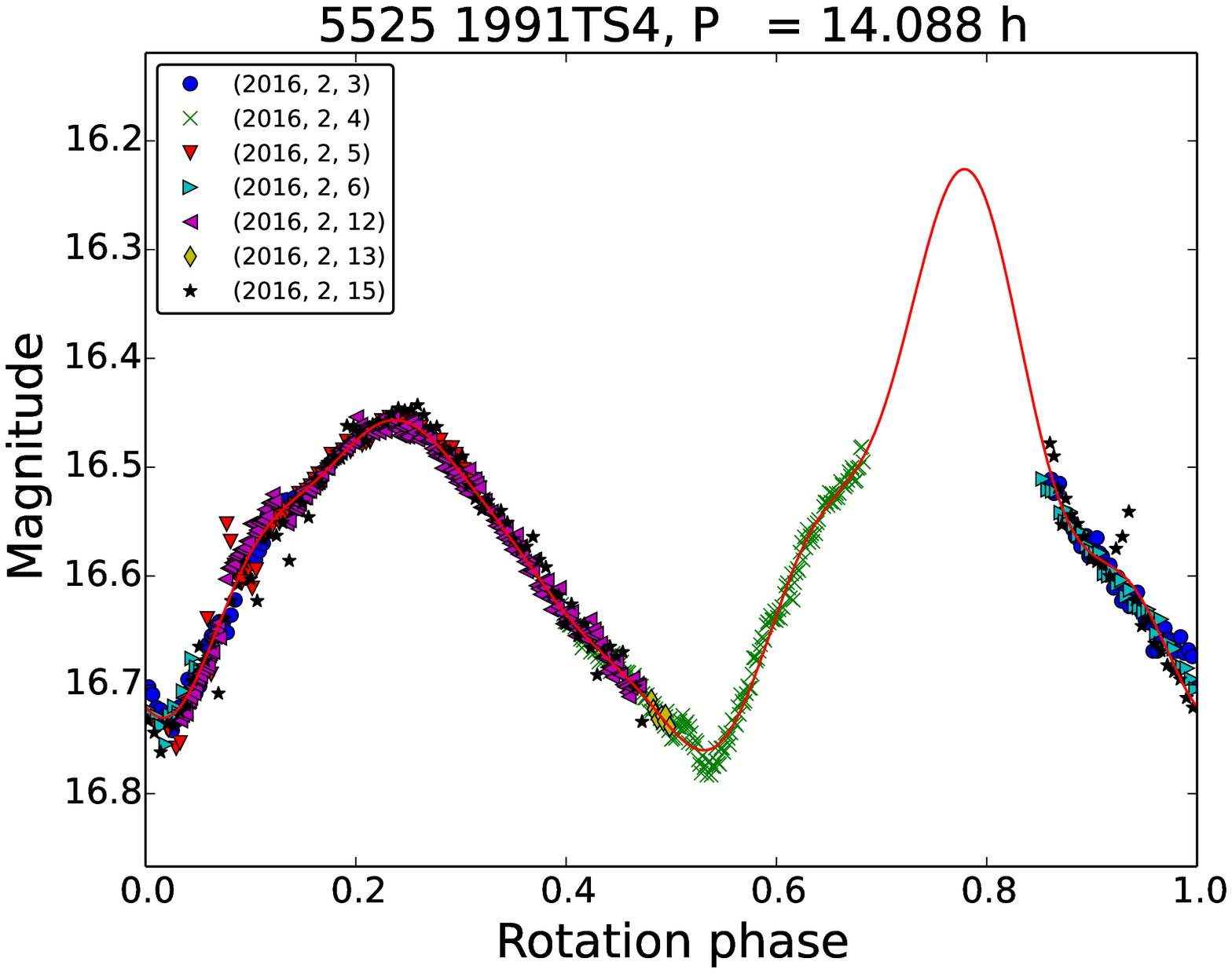}}
    \caption{After opposition. \label{5525at}}
  \end{subfigure} 
    \begin{subfigure}[b]{0.45\linewidth}
    \centering\resizebox{\hsize}{!}{\includegraphics[width=0.5\textwidth]{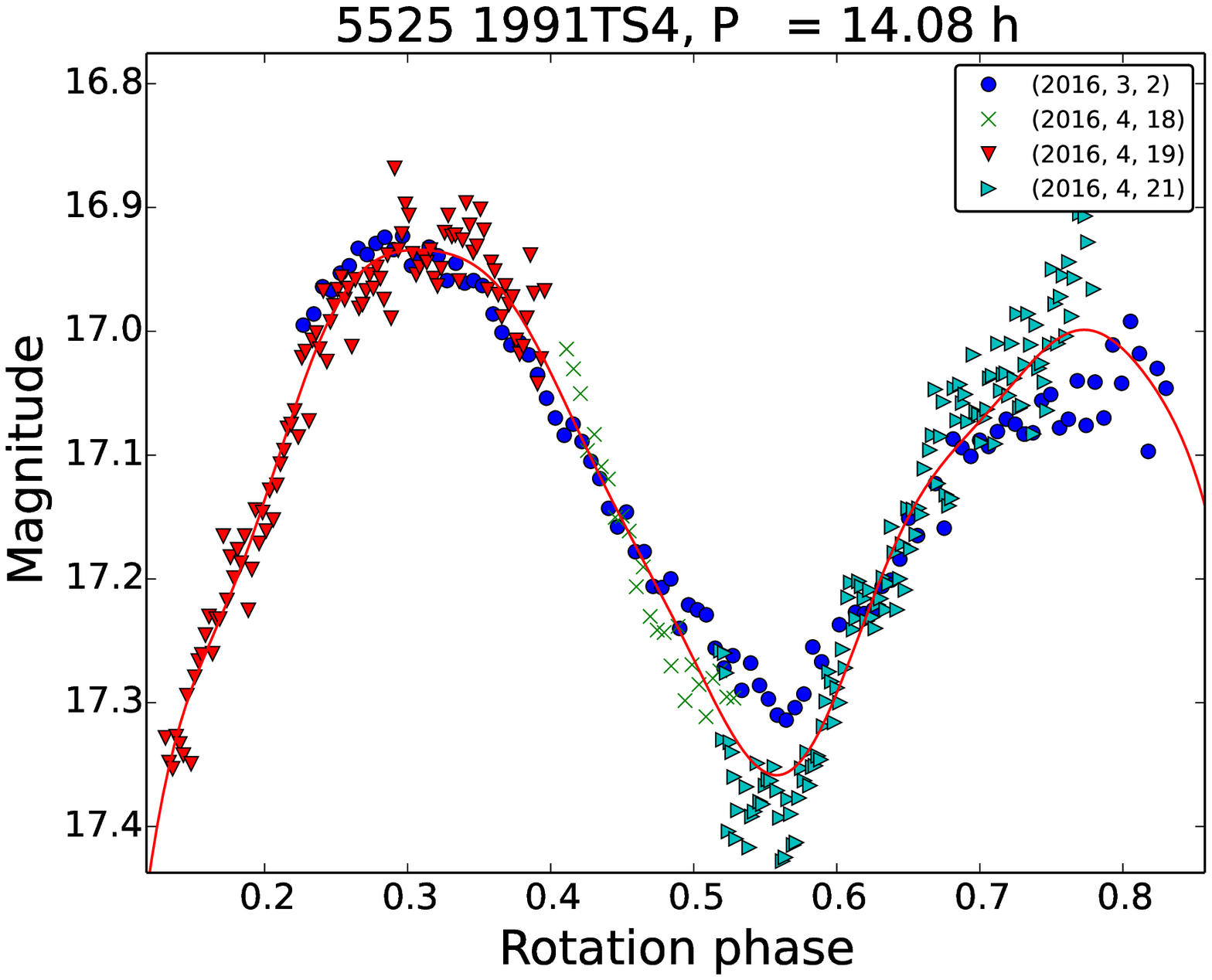}}
    \caption{After opposition. \label{5525a}}
  \end{subfigure}
  \caption{Composite lightcurves for (5525) 1991 TS4. \label{5525lc}}
  \label{5525_lc}
\end{figure}

\begin{figure}[ht]
  \centering
  \begin{subfigure}[b]{0.45\linewidth}
    \centering\resizebox{\hsize}{!}{\includegraphics[width=0.5\textwidth]{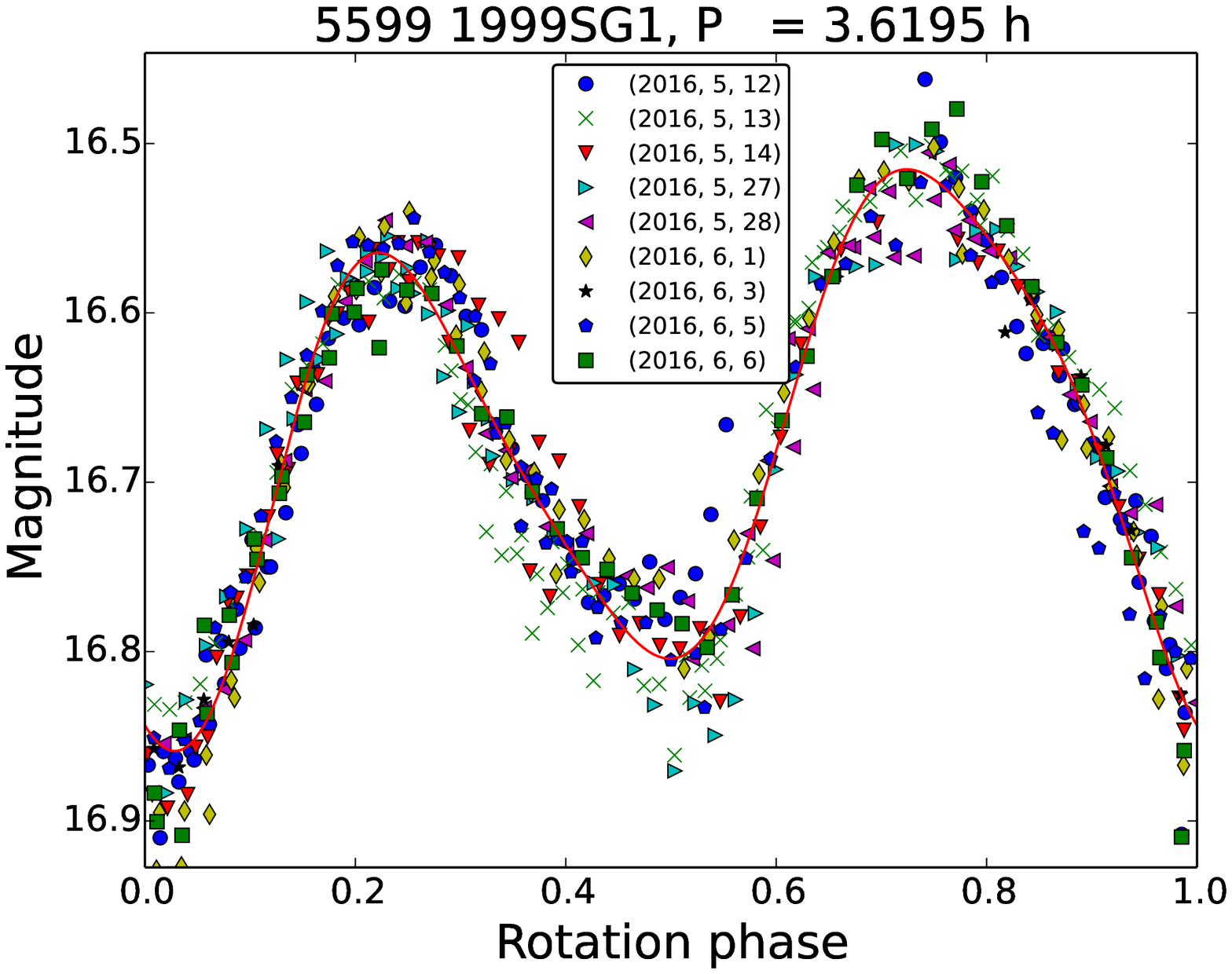}}
    \caption{At opposition. \label{5599Before}}
  \end{subfigure} %
  \begin{subfigure}[b]{0.45\linewidth}
    \centering\resizebox{\hsize}{!}{\includegraphics[width=0.5\textwidth]{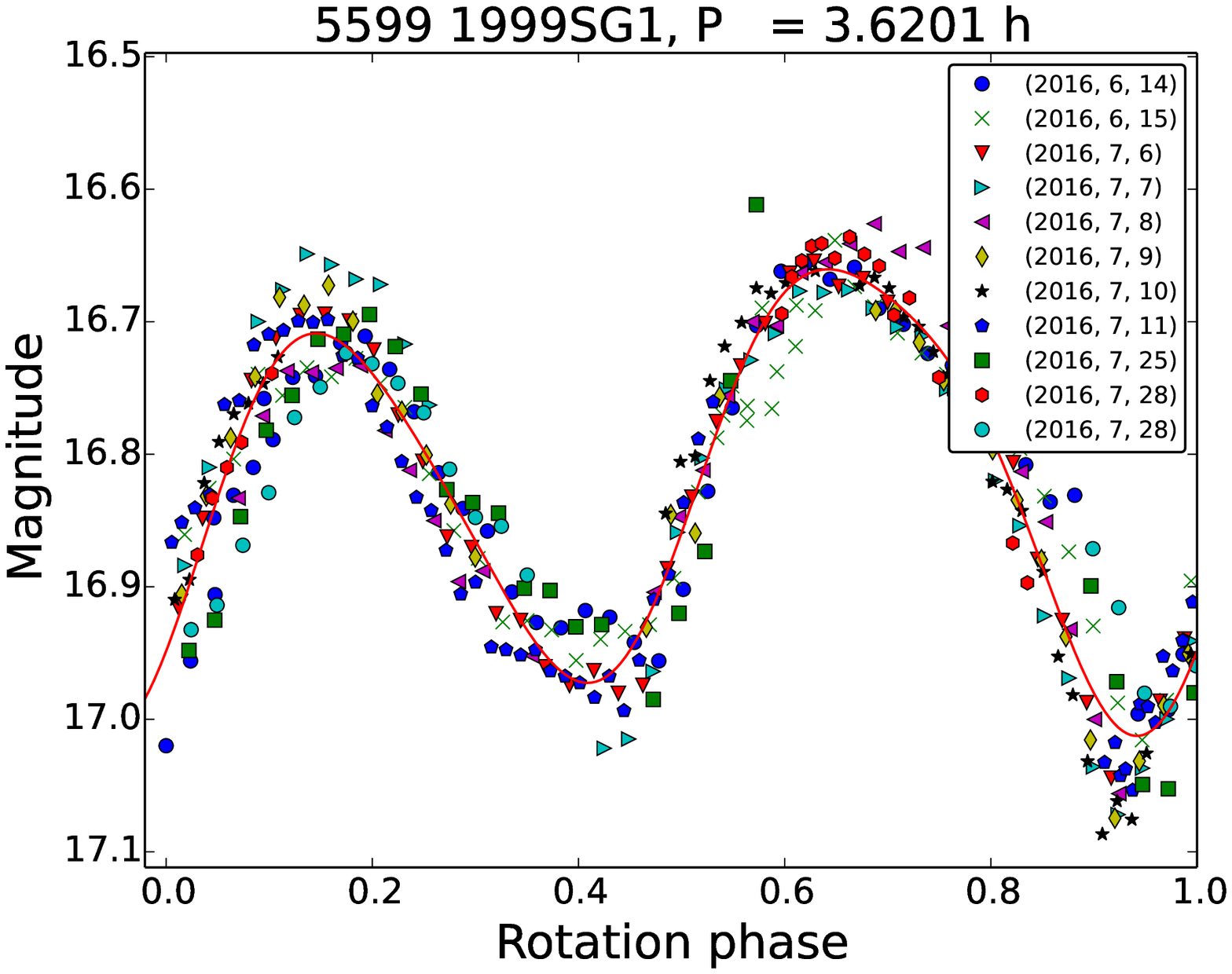}}
    \caption{After opposition. \label{5599At}}
  \end{subfigure}
  \caption{Composite lightcurves for (5599) 1999SG1. \label{5599lc}}
\end{figure}

\begin{figure}[ht]
  \centering
  \begin{subfigure}[b]{0.45\linewidth}
    \centering\resizebox{\hsize}{!}{\includegraphics[width=0.5\textwidth]{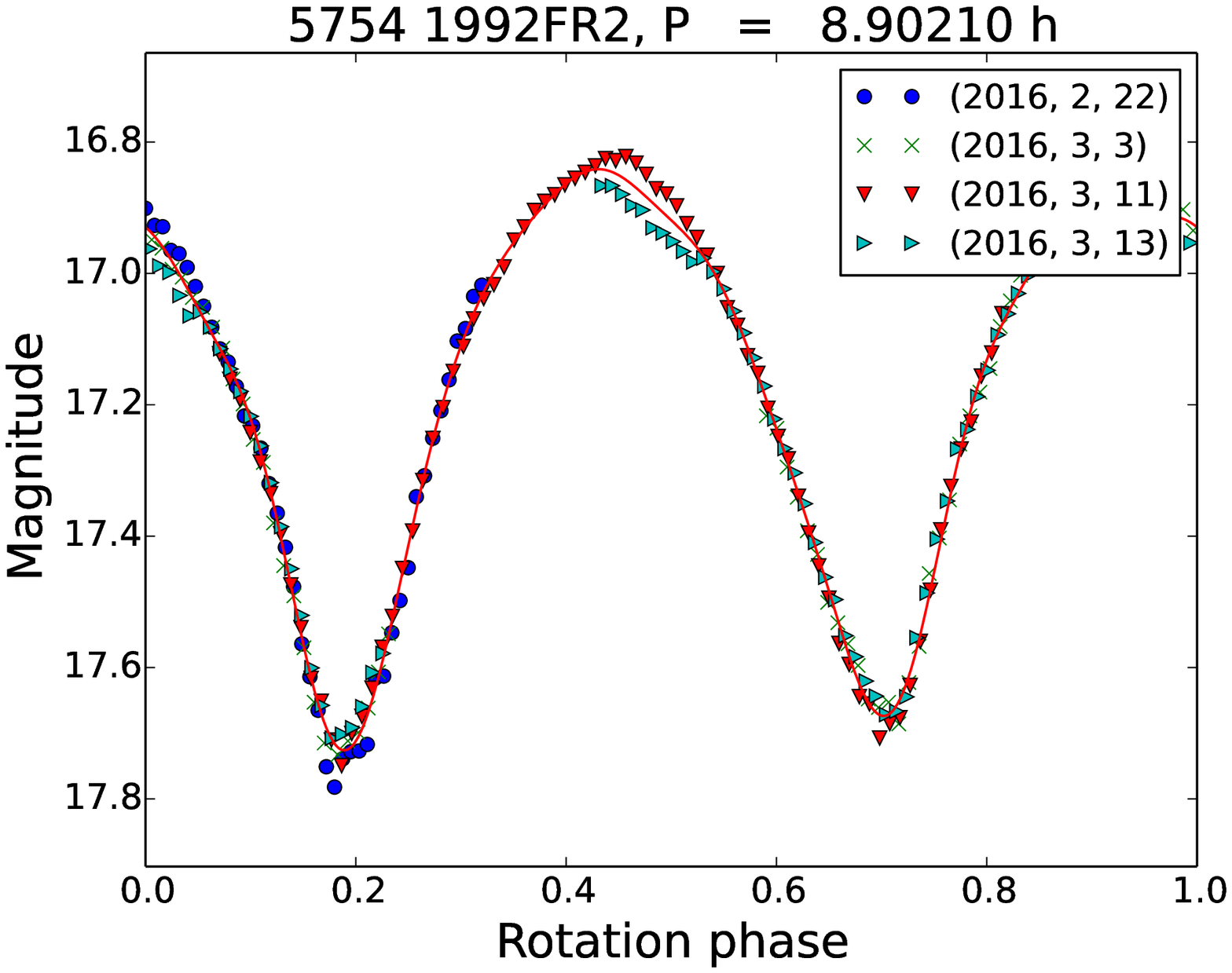}}
    \caption{Before opposition. \label{5754Before}}
  \end{subfigure} %
  \begin{subfigure}[b]{0.45\linewidth}
    \centering\resizebox{\hsize}{!}{\includegraphics[width=0.5\textwidth]{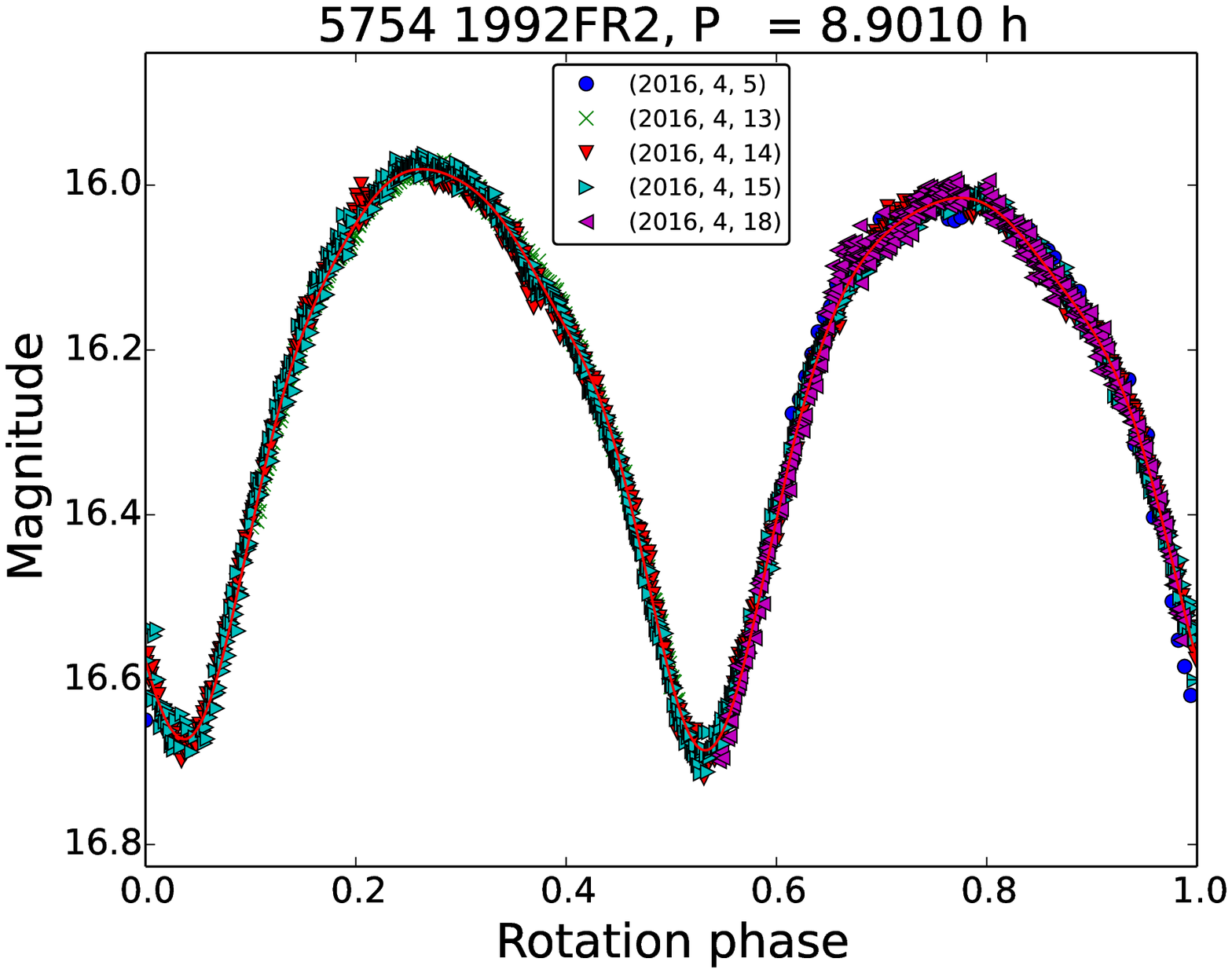}}
    \caption{At opposition. \label{5754At}} 
  \end{subfigure}%
  \\
    \begin{subfigure}[b]{0.45\linewidth}
    \centering\resizebox{\hsize}{!}{\includegraphics[width=0.5\textwidth]{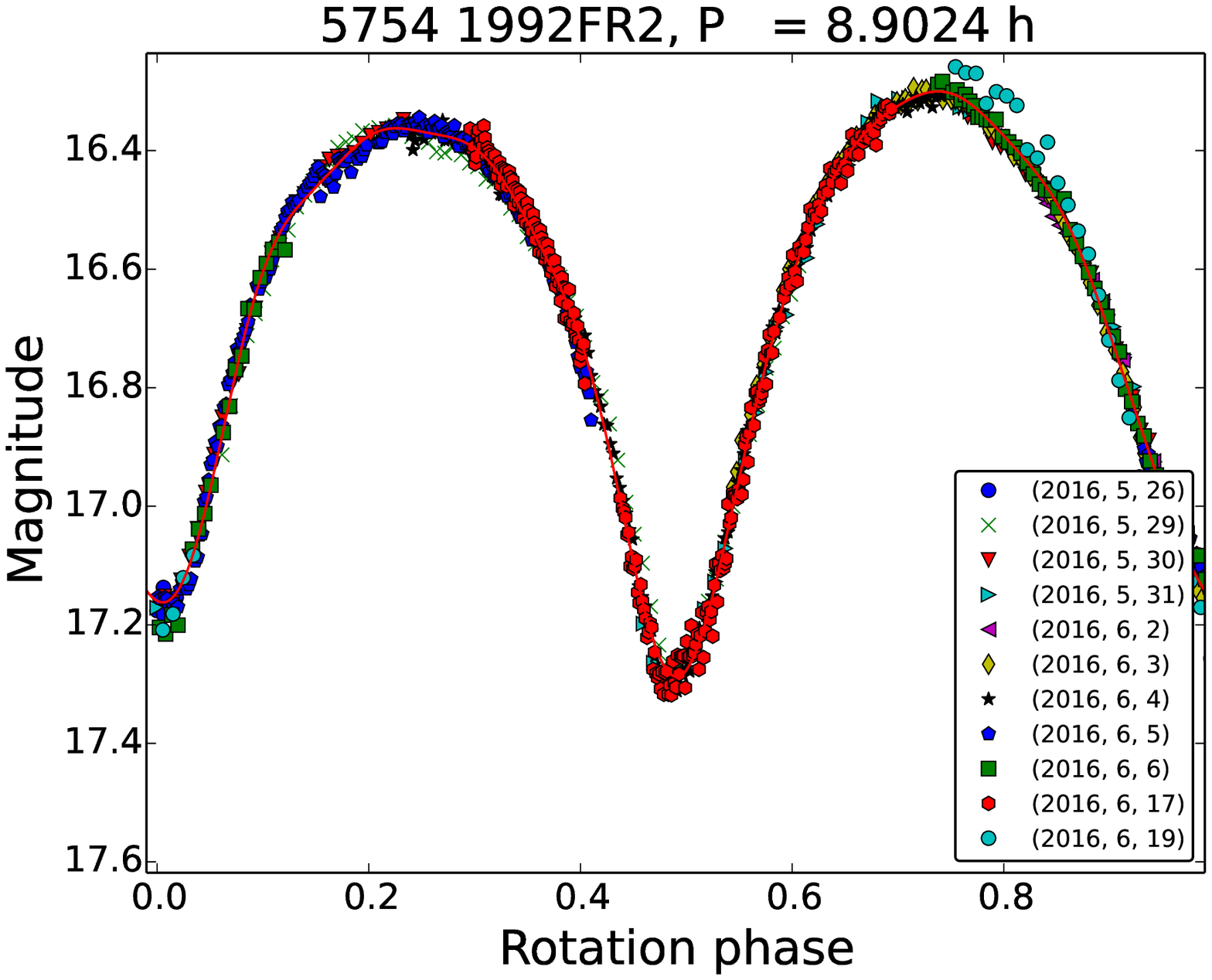}}
    \caption{After opposition. \label{5754Af}}
  \end{subfigure}
  \caption{Composite lightcurves for (5754) 1992 FR2. \label{5754lc}}
\end{figure}

\begin{figure}[ht]
  \centering
  \begin{subfigure}[b]{0.45\linewidth}
    \centering\resizebox{\hsize}{!}{\includegraphics[width=0.5\textwidth]{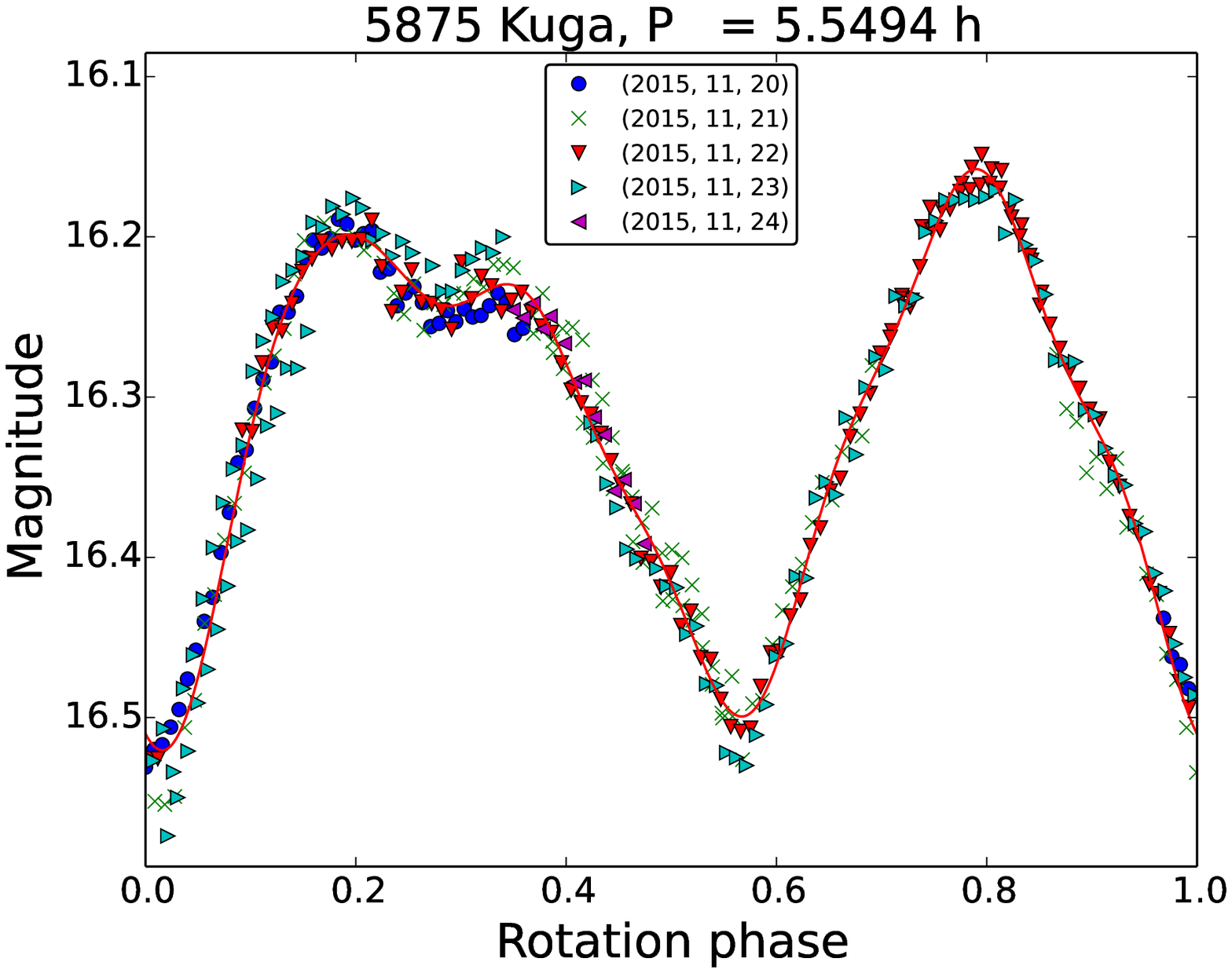}}
    \caption{Before opposition. \label{5857Before}}
  \end{subfigure} %
  \begin{subfigure}[b]{0.45\linewidth}
    \centering\resizebox{\hsize}{!}{\includegraphics[width=0.5\textwidth]{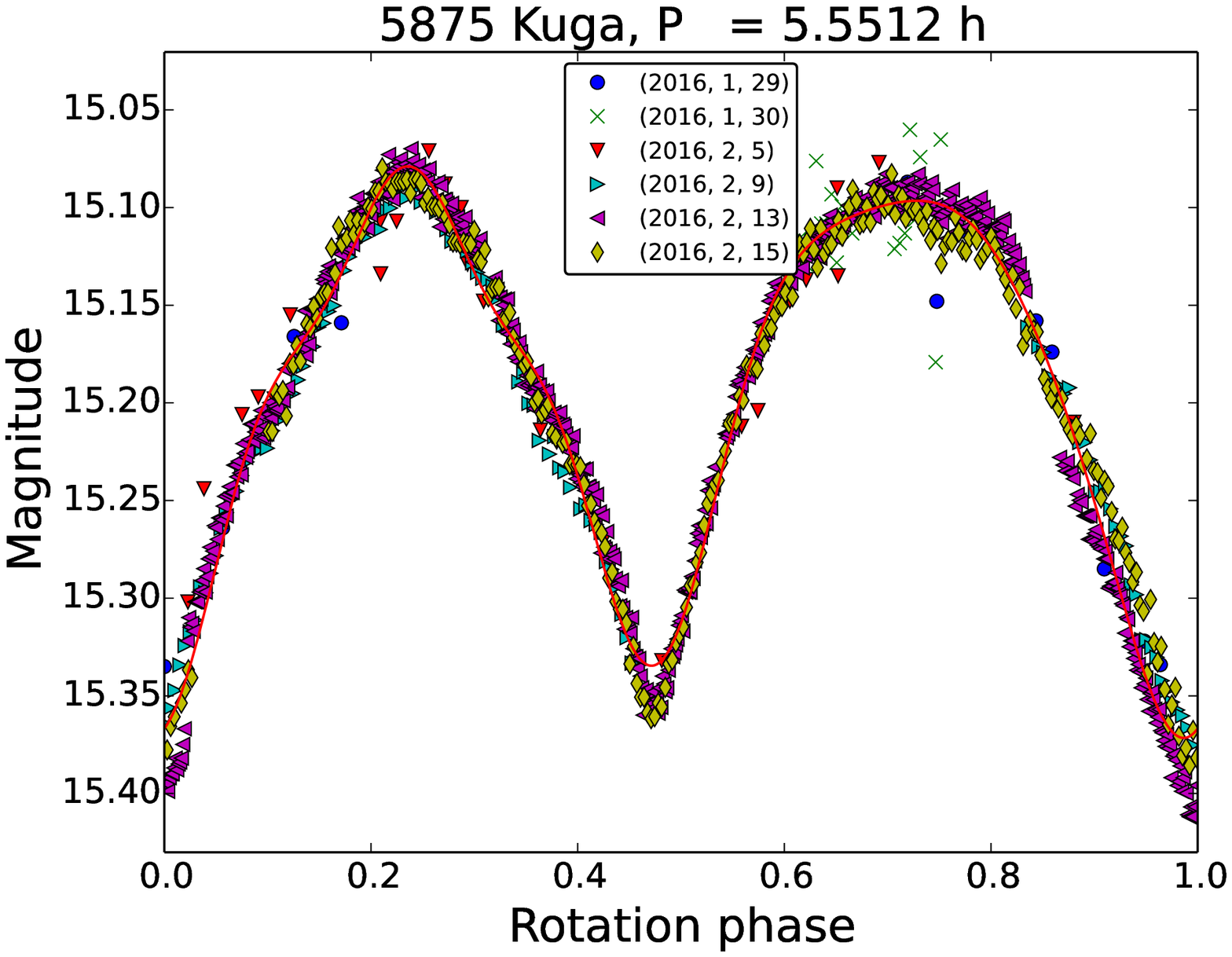}}
    \caption{At opposition. \label{5857At}}
  \end{subfigure}
  \caption{Composite lightcurves for (5857) Kuga. \label{5857lc}}
\end{figure}

\begin{figure}[ht]
  \centering
  \begin{subfigure}[b]{0.45\linewidth}
    \centering\resizebox{\hsize}{!}{\includegraphics[width=0.5\textwidth]{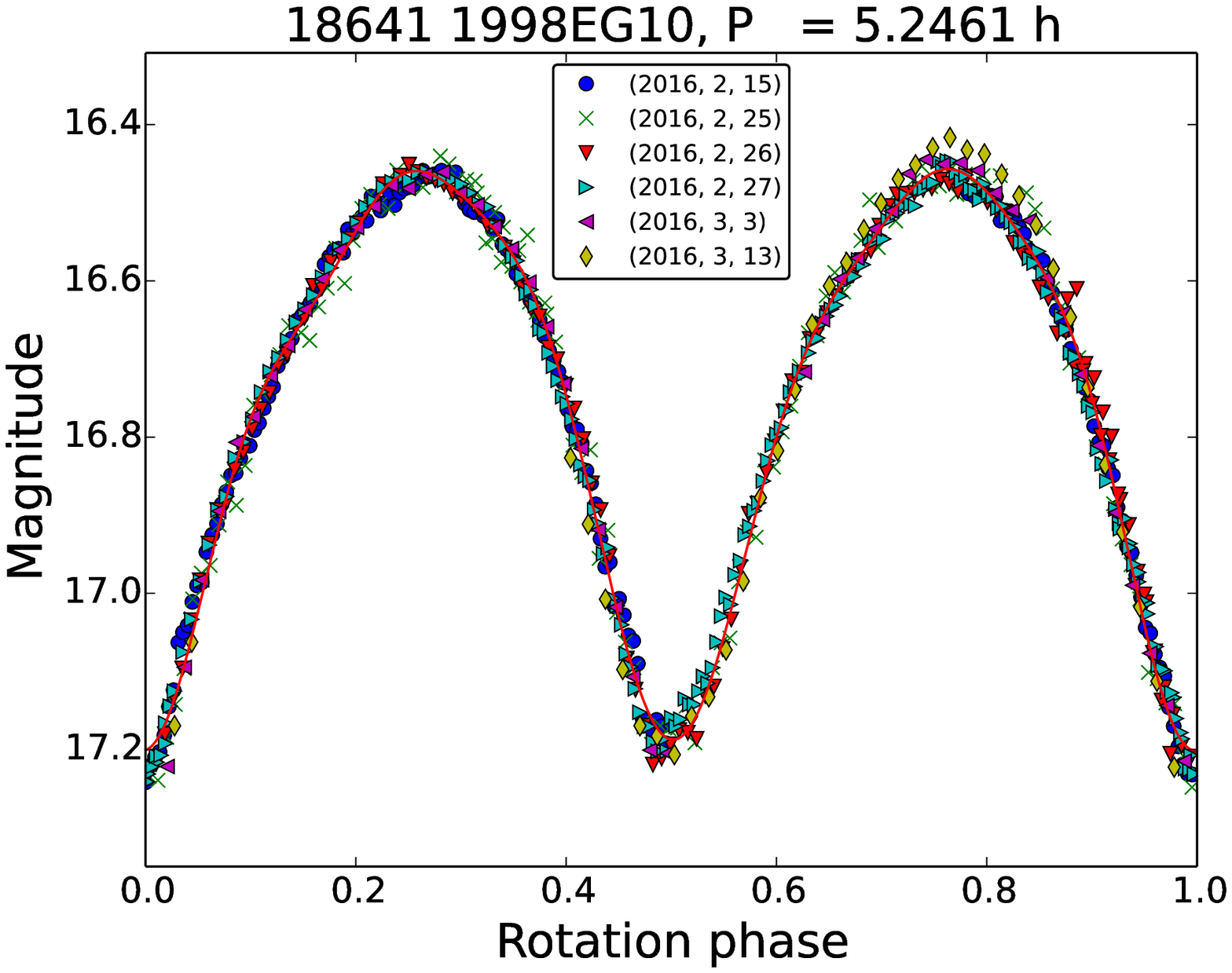}}
    \caption{At opposition. \label{18641At}}
  \end{subfigure} %
  \begin{subfigure}[b]{0.45\linewidth}
    \centering\resizebox{\hsize}{!}{\includegraphics[width=0.5\textwidth]{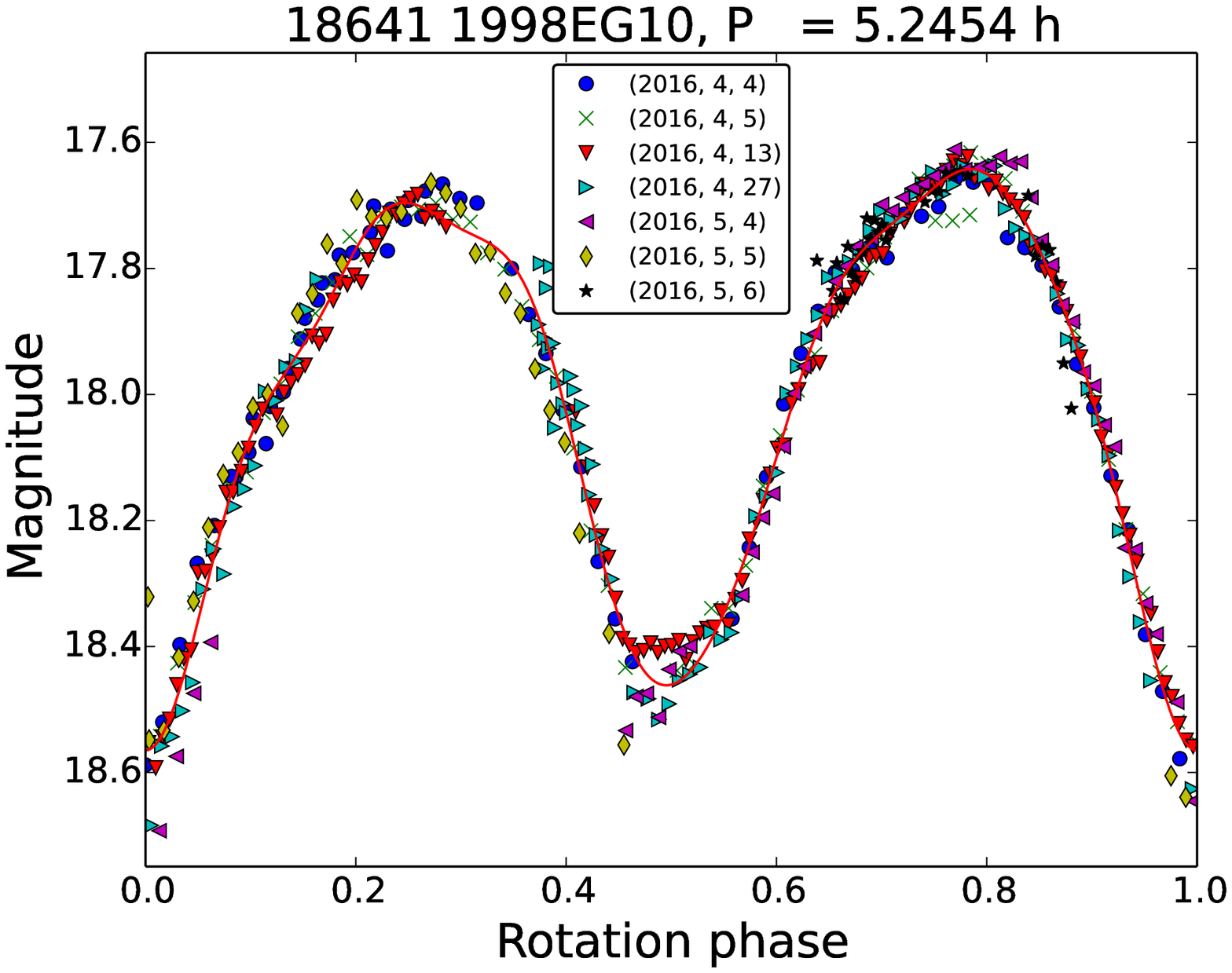}}
    \caption{After opposition. \label{18641After}}
  \end{subfigure}
  \caption{Composite lightcurves for (18641) 1998 EG10. \label{18641lc}}
\end{figure}

\end{appendix}

\end{document}